\documentclass[%
 reprint,
 superscriptaddress,
 amsmath,amssymb,
 aps
]{revtex4-2}

\usepackage{graphicx}
\usepackage{dcolumn}
\usepackage{bm}
\usepackage{amsmath,amsfonts}
\usepackage{latexsym}%
\usepackage{lineno}   
\usepackage{array}
\usepackage{dirtytalk}
\usepackage{csquotes}
\usepackage{adjustbox}
\usepackage{float}
\usepackage{siunitx}
\usepackage{xcolor}
\usepackage[hidelinks]{hyperref}
\usepackage[normalem]{ulem}
 
\begin{document}

\title{\color{black} Resonance behavior of a bubble near a spherical inclusion}
\author{Thomas Micol}
\affiliation{Univ Lyon, Université Claude Bernard Lyon 1, Centre Léon Bérard, INSERM, UMR 1032, LabTAU, F-69003 Lyon, France}
\affiliation{Univ Lyon, INSA Lyon, CNRS, Ecole Centrale de Lyon, Univ Claude Bernard Lyon 1, LMFA, UMR5509, 69621,Villeurbanne France}
\author{Alexander A. Doinikov}
\affiliation{Univ Lyon, Université Claude Bernard Lyon 1, Centre Léon Bérard, INSERM, UMR 1032, LabTAU, F-69003 Lyon, France}
\author{Cyril Mauger}
\affiliation{Univ Lyon, INSA Lyon, CNRS, Ecole Centrale de Lyon, Univ Claude Bernard Lyon 1, LMFA, UMR5509, 69621,Villeurbanne France}
\author{Claude Inserra}
\email{claude.inserra@inserm.fr}
\affiliation{Univ Lyon, Université Claude Bernard Lyon 1, Centre Léon Bérard, INSERM, UMR 1032, LabTAU, F-69003 Lyon, France}
\date{\today} 

\begin{abstract}
We present an analytical model for the frequency response of a gas microbubble oscillating near a spherical inclusion of arbitrary size and mechanical nature (rigid, fluid, or viscoelastic) immersed in a viscous compressible fluid. The model considers both radial and nonspherical oscillations in the linear regime and predicts how their resonance frequencies and oscillation amplitudes are altered by the bubble size, material properties, and distance to the nearby sphere. As a key application, we demonstrate that scanning the frequency response of a bubble near a viscoelastic object, such as an erythrocyte-like particle mimicking a biological cell, offers a way to recover its mechanical properties through inverse modeling, opening new possibilities for high-resolution elastography at the microscale.
\end{abstract}

\maketitle

\section{\label{sec:1} Introduction}
The characterization of the mechanical properties of soft tissues, such as human tissues, plays a fundamental role in the diagnosis and detection of pathological conditions, notably cancer. Indeed, tissues or clusters of cells exhibiting altered mechanical properties often serve as strong indicators of tumor presence and progression \cite{guck2005optical}. Modification of the stiffness and viscoelastic properties of the tissues can reflect variations in cellular architecture, extracellular matrix composition, and pathological remodeling, making mechanical contrast a valuable biomarker in medical imaging. The principal technique that allows the measurement of the mechanical properties of soft tissues is called elastography, which is a relatively recent method first described by \citet{ophir1991elastography}. Amongst the various modalities to assess tissue elasticity, ultrasound elastography provides a non-invasive, real-time assessment of tissue stiffness by tracking shear wave propagation \cite{sarvazyan1998shear}. However, these methods often face limitations in spatial resolution and sensitivity, particularly when probing microscopic structures or complex confined environments, where the key challenge is the generation of elastic waves at a smaller spatial scale than the ultrasound wavelength. \citet{bouchet2024near} showed that acoustically driven microbubbles can be used as an effective means of probing rigid and free interfaces on a millimeter scale, offering alternative solutions to traditional elastography methods mentioned above.

Gas microbubbles exposed to ultrasonic fields act as high-quality acoustic resonators, capable of undergoing strong radial oscillations and, under certain conditions, inertial cavitation \cite{plesset1977bubble}. This phenomenon has been widely exploited in applications such as ultrasonic cleaning, where oscillating bubbles remove contaminants from surfaces, and in medical procedures like lithotripsy, where bubble collapses help in fragmenting kidney stones \cite{leighton1994acoustic, johnsen2008shock}. Beyond these traditional applications, microbubbles are also employed in biomedical imaging as ultrasound contrast agents due to their strong acoustic scattering properties \cite{hoff2001acoustic}, and in targeted drug delivery, where their collapse can transiently permeabilize cell membranes \cite{ferrara2007ultrasound}.

At low acoustic amplitudes, microbubbles undergo stable cavitation, oscillating steadily about their equilibrium radius without collapsing. In this regime, the bubble behaves as a linear acoustic resonator with a well-defined resonance frequency (the so-called Minnaert frequency), determined by its size and the properties of the surrounding medium. This resonance is highly sensitive to its environment \cite{dollet2019bubble, vincent2017statics}, making it a powerful probe for characterizing local mechanical properties.

The effect of nearby boundaries on the bubble resonance has been the subject of extensive experimental and theoretical work. There exist several experimental studies in literature dealing with bubbles close to or attached to rigid, elastic walls, or free interfaces \cite{payne2005symmetric, garbin2007changes, helfield2014effect, baresch2020acoustic, bouchet2024near}. These studies have highlighted the complex effect of confinement on the acoustic response of the bubble, and comparisons with theoretical models are not always feasible due to the lack of versatility of existing models.

The case of a plane, rigid or free interface has been considered by \citet{strasberg1953pulsation} by introducing a mirror bubble. It has been shown that a nearby rigid interface lowers the bubble’s natural frequency, whereas a free interface raises it. This approach, however, is limited to inviscid fluids and large bubble-wall distances. \citet{leighton1994acoustic} extended this model by incorporating multiple scattering effects between the bubble and its image. Doinikov later developed more rigorous formulations based on multipole expansions, allowing for arbitrary bubble-wall distances and introducing viscoelastic boundaries \cite{doinikov2009modeling, doinikov2011acoustic, doinikov2011dynamics, doinikov2015interaction}. The case of viscous and compressible fluids has been investigated by \citet{hay2012model} for a bubble oscillating near one or between two viscoelastic walls of finite thickness. This modeling allows the derivation of the frequency response of the bubble at arbitrary distance to the wall. The dynamics of non spherically-oscillating microbubbles near rigid walls was investigated theoretically by \citet{maksimov2020splitting} and using boundary integral methods by \citet{corbett2023numerical}. Recently, \citet{liu2022natural} have considered the interaction between an oscillating bubble and a finite-size rigid sphere in an incompressible fluid at low viscosity, also including shape modes. In their paper, they reported that \enquote{natural frequencies of the bubble are decreased or even eliminated \ldots as the solid sphere gets bigger and closer}. This conclusion, however, appears to be difficult to reconcile with most of the existing literature, where no evidence has been found for the disappearance of bubble oscillations near a rigid interface.

In summary, a unified theoretical framework capable of predicting the modal frequency response of a bubble near a curved interface of arbitrary size and material (rigid, fluid, or viscoelastic) remains lacking. In this work, we address this gap by developing an analytical model for a bubble oscillating near a sphere, assuming a viscous and compressible surrounding fluid. The model considers both the radial and non-spherical oscillations, providing a versatile tool to investigate how the resonance characteristics of the bubble are modified by the mechanical nature, size, and proximity of the neighboring object. In Sec.~\ref{sec:2}, the linear acoustic velocity field in the fluid surrounding the bubble and the viscoelastic particle is derived, which allows the prediction of the bubble oscillation amplitude and frequency response. Applications of the developed modeling are investigated numerically in Sec.~\ref{sec:3}.

\section{\label{sec:2} Theory}
\begingroup
\allowdisplaybreaks
To analyze the influence of a nearby object on the resonance characteristics of a gas bubble, we consider a minimal yet representative configuration: a spherical gas bubble of equilibrium radius $R_{10}$, located at a distance $d$ from the center of a sphere of radius $R_{20}$, both immersed in a viscous compressible liquid. This geometry provides a tractable model for studying the acoustic coupling between an oscillating bubble and a close spherical body.

Both the bubble and the sphere are assumed to be spherical at rest, centered at the origin of their respective coordinate systems $(r_1, \theta_1, \varepsilon)$ and $(r_2, \theta_2, \varepsilon)$, as shown in Fig.~\ref{fig:FIG1}. The geometry is considered as axisymmetric with respect to the azimuthal angle $\varepsilon$.

\begin{figure}[t]
    \centering
    \includegraphics[width=\linewidth]{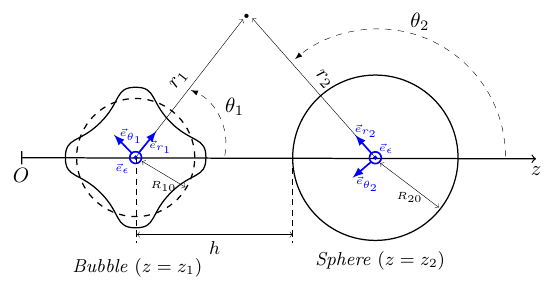}
    \caption{Coordinate systems used in the theoretical model. The system is considered as axisymmetric around the $z-$axis.}
    \label{fig:FIG1}
\end{figure}

The bubble is subjected to small amplitude oscillations at angular frequency $\omega$, and its interface can exhibit a set of surface modes indexed by their degrees $m \in \{M_1, ..., M_N\}$. The perturbed interface of the bubble is described by
\begin{equation}
    R_1(\mu_1, t) = R_{10} + \mathrm{e}^{-i \omega t} \sum_{m = M_1}^{M_N} s_m P_m(\mu_1),
    \label{eq:1}
\end{equation}
where $\mu_1 = \cos \theta_1$, $s_m$ is the complex amplitude of the $m$th mode, and $P_m$ is the Legendre polynomial of degree $m$.

In the first part of the analysis, the modal amplitudes $s_m$ are considered known, for instance, obtained experimentally. In the second part, they are computed from a theoretical model that relates them to the amplitude of the driving acoustic pressure and to the geometric and physical parameters of the system.

\subsection{\label{sec:2a} First-order acoustic field in the liquid}
The linearized first-order equations of motion of a viscous compressible liquid are given by \citet{landau1987fluid},
\begin{gather}
    \frac{\partial\rho}{\partial t} + \rho_0\boldsymbol{\nabla}\cdot \boldsymbol{v}=0,\label{eq:2b}\\
    \rho_0\frac{\partial \boldsymbol{v}}{\partial t}=-\boldsymbol{\nabla}p+\eta\Delta\boldsymbol{v}+\left(\xi + \frac{\eta}{3}\right)\boldsymbol{\nabla}\left(\boldsymbol{\nabla}\cdot\boldsymbol{v}\right),\label{eq:3b}\\
    p=c^2\rho\label{eq:4b},
\end{gather}
where $\rho$ is the first-order liquid density, $\rho_0$ is the equilibrium liquid density, $\boldsymbol{v}$ is the first-order liquid velocity, $p$ is the first-order liquid pressure, $\eta$ is the dynamic viscosity, $\xi$ is the volume viscosity, and $c$ is the speed of sound.

The first-order velocity $\boldsymbol{v}$ is written as
\begin{equation}
    \boldsymbol{v}=\boldsymbol{v}^{(1)} + \boldsymbol{v}^{(2)}, 
\end{equation}
where $\boldsymbol{v}^{(1)}$ and $\boldsymbol{v}^{(2)}$ are the velocities generated by the bubble and the sphere, respectively.

Following the Helmholtz decomposition, the solution of Eqs.~(\ref{eq:2b}) -- (\ref{eq:4b}) is sought as 
\begin{equation}
    \boldsymbol{v}^{(j)}=\boldsymbol{\nabla}\varphi^{(j)} + \boldsymbol{\nabla}\times\boldsymbol{\psi}^{(j)},\label{eq:5b}
\end{equation}
where $\varphi$ and $\boldsymbol{\psi}$ are the scalar and the vector velocity potentials, respectively, and $j=1,2$.

Substituting Eq.~(\ref{eq:5b}) into Eqs.~(\ref{eq:2b}) -- (\ref{eq:4b}) and considering the harmonic ($\mathrm{e}^{-i\omega t}$) time dependence, one obtains that the scalar and vector potentials should obey the following equations:
\begin{gather}
    \Delta\varphi^{(j)} + k_a^2\varphi^{(j)} = 0,\label{eq:7b}\\
    \Delta\boldsymbol{\psi}^{(j)} + k_v^2\boldsymbol{\psi}^{(j)} = 0\label{eq:8b},
\end{gather}
where
\begin{gather}
k_a=\frac{\omega}{c}\left[1-\frac{i \omega}{\rho_0 c^2}\left(\xi+\frac{4 \eta}{3}\right)\right]^{-\frac{1}{2}}, \label{eq:4}\\
k_v=\frac{1+i}{\delta}, \quad \delta=\sqrt{\frac{2\nu}{\omega}}.\label{eq:5}
\end{gather}
Here, $k_a$ is the acoustic wavenumber and $k_v$ is the viscous wavenumber, where $\nu=\eta/\rho_0$ is the kinematic viscosity and $\delta$ is the viscous penetration depth.

In view of the axial symmetry of the problem, solutions to Eqs.~(\ref{eq:7b}) and (\ref{eq:8b}) give the velocity potentials of the waves generated by the bubble and the sphere (wave 1 or wave 2, respectively) in the coordinates of the bubble $\left(r_1, \theta_1\right)$ and in the coordinates of the sphere $\left(r_2, \theta_2\right)$ as
\begin{align}
\varphi^{(j)}\left(r_j, \theta_j, t\right)&=\mathrm{e}^{-i \omega t} \sum_{n=0}^{\infty} a_n^{(j)} h_n^{(1)}\left(k_a r_j\right) P_n\left(\mu_j\right),\label{eq:2} \\
\boldsymbol{\psi}^{(j)}\left(r_j, \theta_j, t\right)&=\mathrm{e}^{-i \omega t} \boldsymbol{e}_{\varepsilon} \sum_{n=1}^{\infty} b_n^{(j)} h_n^{(1)}\left(k_v r_j\right) P_n^1\left(\mu_j\right),\label{eq:3}
\end{align}
where $\mu_j=\cos\theta_j$, $P_n^1$ is the associated Legendre polynomial of the first order and degree $n$, and $\boldsymbol{e}_\varepsilon$ is the azimuthal vector represented in Fig.~\ref{fig:FIG1}. The function $h_n^{(1)}(x)$ is the spherical Hankel function of the first kind, representing an outgoing wave. Note, however, that the general solutions of Eqs.~(\ref{eq:7b}) and (\ref{eq:8b}) also involve the function $h_n^{(2)}(x)$, which represents an incoming wave and is not relevant in the present situation. Note also that calculating Eqs.~(\ref{eq:2}) and (\ref{eq:3}), we have used the mathematical properties of $P_n(\mu)$ and $P_n^1(\mu)$ \citep{arfken1985mathematical, abramowitz1965handbook}.

Substituting Eqs.~(\ref{eq:2}) and (\ref{eq:3}) into Eq.~(\ref{eq:5b}) yields the velocity components of the waves generated by the bubble and the sphere,
\begin{align}
v_r^{(j)}\left(r_j, \theta_j, t\right)&=\frac{\mathrm{e}^{-i \omega t}}{r_j} \sum_{n=0}^{\infty}\left[a_n^{(j)} k_a r_j h_n^{(1) '}\left(k_a r_j\right)\right.\nonumber\\*
&\left.-b_n^{(j)} n(n+1) h_n^{(1)}\left(k_v r_j\right)\right] P_n\left(\mu_j\right),\label{eq:6} \\
v_\theta^{(j)}\left(r_j, \theta_j, t\right)&=\frac{\mathrm{e}^{-i \omega t}}{r_j} \sum_{n=1}^{\infty}\left\{a_n^{(j)} h_n^{(1)}\left(k_a r_j\right)\right.\nonumber\\*
&\left.-b_n^{(j)}\left[h_n^{(1)}\left(k_v r_j\right)+k_v r_j h_n^{(1) '}\left(k_v r_j\right)\right]\right\}\nonumber\\*
&\times P_n^1\left(\mu_j\right) .\label{eq:7}
\end{align}
In order to find the linear scattering coefficients $a_n^{(j)}$ and $b_n^{(j)}$, we need to express the boundary conditions at the sphere's and at the bubble's surface. Hence, the velocity components of the wave generated by the bubble (respectively by the sphere) must be expressed in the coordinate system of the sphere $(r_2, \theta_2)$ (respectively of the bubble $(r_1, \theta_1)$). These coordinate transformations are provided in Appendix~\ref{sec:app1}.

The first boundary condition for the liquid velocity at the surface of the bubble requires that the normal component of $\boldsymbol{v}$ be equal to the normal component of the surface velocity of the bubble. The second condition requires that the tangential stress vanish because the gas viscosity is much lower than the liquid viscosity. It is worth noticing that, for applications using ultrasound contrast agents, the lipid or polymer shell introduces surface shear viscosity and elasticity, which alters this boundary condition towards a no-slip or complex stress-balance behavior, which is beyond the scope of the present study. These boundary conditions are written by
\begin{gather}
    v_r=\frac{\partial R_1}{\partial t}=-i\omega \mathrm{e}^{-i\omega t}\sum_{m=M_1}^{M_N}s_mP_m(\mu_1),\label{eq:28}\\
    \sigma_{r\theta}=0,\label{eq:29}
\end{gather}
at $r_1=R_{10}$. The radial component of the velocity field $v_r=v_r^{(1)}+v_r^{(2)}$ is calculated in the coordinate system of the bubble $(r_1,\theta_1)$ by using the velocity radiated by the bubble (Eq.~(\ref{eq:6}) with $j=1$) and the one radiated by the sphere, which is written in the same coordinate system (given by Eq.~(\ref{eq:24})). Similarly, the tangential component of the velocity field $v_\theta=v_\theta^{(1)} + v_\theta^{(2)}$ is calculated using Eq.~(\ref{eq:7}) with $j=1$ and Eq.~(\ref{eq:25}). The tangential stress $\sigma_{r\theta}$ is given by \citet{landau1987fluid},
\begin{equation}
    \sigma_{r\theta}=\eta\left(\frac{1}{r_1}\frac{\partial v_r}{\partial\theta_1}+\frac{\partial v_\theta}{\partial r_1}-\frac{v_\theta}{r_1}\right).\label{eq:sigma_rt}
\end{equation}
These boundary conditions are calculated in Appendix~\ref{sec:app21}.

In Secs.~\ref{sec:2a1}, \ref{sec:2a2}, and \ref{sec:2a3}, the boundary conditions at the surface of the spherical particle are derived and combined with the boundary conditions at the bubble interface, Eqs.~(\ref{eq:36}) and (\ref{eq:37}), in order to determine the linear scattering coefficients $a_n^{(j)}$ and $b_n^{(j)}$. Three configurations are considered: a perfectly rigid sphere, a viscous compressible fluid sphere, and a viscoelastic solid sphere. In the latter two cases, the wave propagation inside the sphere must also be taken into account, introducing two additional scattering coefficients, $\hat{a}_n$ and $\hat{b}_n$, and requiring the application of two supplementary boundary conditions at the sphere's surface.

\subsubsection{\label{sec:2a1} Case of a rigid sphere}
The boundary conditions on the surface of a rigid sphere require that both the normal and tangential components of $\boldsymbol{v}$ vanish. These conditions can be written by
\begin{gather}
    v_r= 0,\label{eq:32}\\
    v_\theta = 0,\label{eq:33}
\end{gather}
at $r_2=R_{20}$. The radial component of the velocity field $v_r=v_r^{(1)}+v_r^{(2)}$ is calculated in the coordinate system of the sphere $(r_2,\theta_2)$ using the velocity radiated by the sphere (Eq.~(\ref{eq:6}) with $j=2$) and the one radiated by the bubble, which is written in the same coordinate system (given by Eqs.~(\ref{eq:12})). Similarly, the tangential component of the velocity field $v_\theta=v_\theta^{(1)} + v_\theta^{(2)}$ is calculated using Eq.~(\ref{eq:7}) with $j=2$ and Eq.~(\ref{eq:13}).

These boundary conditions are calculated in Appendix~\ref{sec:app22}. Combining the resulting equations, given by Eqs.~(\ref{eq:39}) and (\ref{eq:54b}), with the boundary conditions at the surface of the bubble given by Eqs.~(\ref{eq:36}) and (\ref{eq:37}), we get a system of four equations, which can be solved numerically in order to find the linear scattering coefficients $a_n^{(1)}$, $a_n^{(2)}$, $b_n^{(1)}$ and $b_n^{(2)}$.

\subsubsection{\label{sec:2a2} Case of a viscous compressible fluid sphere}

We now consider a viscous compressible fluid sphere centered at the origin of the coordinates system $(r_2,\theta_2)$, with density $\rho_s$, sound speed $c_s$, volume fluid viscosity $\xi_s$ and dynamic viscosity $\eta_s$.

The velocity potentials of the wave inside the sphere can be written by
\begin{align}
\hat{\varphi}(r_2, \theta_2, t)&=\mathrm{e}^{-i \omega t} \sum_{n=0}^{\infty} \hat{a}_n j_n(\hat{k}_a r_2) P_n(\mu_2),\label{eq:67} \\
\hat{\psi}(r_2, \theta_2, t)&=\mathrm{e}^{-i \omega t} \boldsymbol{e}_{\varepsilon} \sum_{n=1}^{\infty} \hat{b}_n j_n(\hat{k}_v r_2) P_n^1(\mu_2),\label{eq:68}
\end{align}
thus leading to the following velocity components:
\begin{align}
\hat{v}_r(r_2,\theta_2,t)&=\frac{\mathrm{e}^{-i\omega t}}{r_2}\sum_{n=0}^\infty \left[\hat{a}_n\hat{k}_ar_2j_n'(\hat{k}_ar_2)\right.\nonumber\\*
&\left.-\hat{b}_nn(n+1)j_n(\hat{k}_vr_2)\right]P_n(\mu_2),\\
\hat{v}_\theta(r_2,\theta_2,t)&=\frac{\mathrm{e}^{-i\omega t}}{r_2}\sum_{n=1}^\infty \left\{\hat{a}_nj_n(\hat{k}_ar_2) \right.\nonumber\\*
&\left.-\hat{b}_n\left[j_n(\hat{k}_v r_2) + \hat{k}_v r_2j_n'(\hat{k}_vr_2)\right]\right\}\nonumber\\*
&\times P_n^1(\mu_2),
\end{align}
where $j_n(x)$ denotes the spherical Bessel function of the first kind. Although the general solution of Eqs.~(\ref{eq:67}) and (\ref{eq:68}) also includes the spherical Bessel function of the second kind $y_n(x)$, this term is discarded in the present case since $y_n(x)$ diverges at $x = 0$, whereas the physical solution is required to remain finite at the center of the sphere.

The boundary conditions at the surface of a viscous compressible fluid sphere require the continuity of the normal and tangential components of $\boldsymbol{v}$, as well as the continuity of the normal and tangential stress. These conditions can be written by
\begin{gather}
    v_r=\hat{v}_r,\label{eq:71}\\
    v_\theta=\hat{v}_\theta,\label{eq:72} \\
    \sigma_{r\theta}=\hat{\sigma}_{r\theta},\label{eq:73}\\
    \sigma_{rr}=\hat{\sigma}_{rr},\label{eq:74}
\end{gather}
at $r_2=R_{20}$, where $\sigma_{rr}$ and $\sigma_{r\theta}$ are the first-order normal and tangential stresses in the liquid and $\hat{\sigma}_{rr}$ and $\hat{\sigma}_{r\theta}$ are the first-order normal and tangential stresses in the sphere. The tangential stress $\sigma_{r\theta}$ is given by Eq.~(\ref{eq:sigma_rt}), while $\hat{\sigma}_{r\theta}$ is given by the same mathematical expression, replacing $\eta$, $v_r$ and $v_\theta$ by $\eta_s$, $\hat{v}_r$ and $\hat{v}_\theta$. The normal stress $\sigma_{rr}$ is given by
\begin{equation}
    \sigma_{rr}=-p + 2\eta\frac{\partial v_r}{\partial r_1}+\left(\xi - \frac{2}{3}\eta\right)\boldsymbol{\nabla}\cdot\boldsymbol{v},\label{eq:72b}
\end{equation}
where $\boldsymbol{v}=\boldsymbol{v}^{(1)}+\boldsymbol{v}^{(2)}$ and $p=p^{(1)} + p^{(2)}$. The contributions $p^{(1)}$ and $p^{(2)}$ are the pressures generated by the bubble and the sphere, respectively. It follows from Eqs.~(\ref{eq:2b}), (\ref{eq:4b}) and (\ref{eq:7b}) that
\begin{equation}
    p^{(j)}=\frac{i\rho_0 c^2 k_a^2}{\omega}\varphi^{(j)}.
\end{equation}
The normal stress $\hat{\sigma}_{rr}$ in the sphere is given by the same mathematical expression as $\sigma_{rr}$, replacing $\eta$, $v_r$ and $\boldsymbol{v}$ by $\eta_s$, $\hat{v}_r$ and $\boldsymbol{\hat{v}}$.

The boundary conditions given by Eqs.~(\ref{eq:71}) -- (\ref{eq:74}) are calculated in Appendix~\ref{sec:app23}. Combining the resulting equations, given by Eqs.~(\ref{eq:81}) -- (\ref{eq:84}), with the boundary conditions at the surface of the bubble given by Eqs.~(\ref{eq:36}) and (\ref{eq:37}), a system of six equations is obtained, which can be solved numerically in order to find the linear scattering coefficients $a_n^{(1)}$, $a_n^{(2)}$, $b_n^{(1)}$, $b_n^{(2)}$, $\hat{a}_n$ and $\hat{b}_n$.

\subsubsection{\label{sec:2a3} Case of a solid viscoelastic sphere}
We now consider a solid viscoelastic sphere centered at the origin of the coordinates system $(r_2,\theta_2)$, and we assume that the motion of the viscoelastic medium inside the sphere obeys the following equation \citep{landau1970theory}:
\begin{align}
    \rho_p\frac{\partial^2 \boldsymbol{u}}{\partial t^2}&=\mu_p\Delta\boldsymbol{u}+(\lambda_p+\mu_p)\boldsymbol{\nabla}(\boldsymbol{\nabla}\cdot \boldsymbol{u}) \nonumber\\*
    &+ \eta_p\Delta\frac{\partial \boldsymbol{u}}{\partial t}+\left(\xi_p + \frac{1}{3}\eta_p\right)\boldsymbol{\nabla}\left(\boldsymbol{\nabla}\cdot\frac{\partial \boldsymbol{u}}{\partial t}\right),
    \label{eq:109}
\end{align}
where $\boldsymbol{u}$ is the displacement vector, $\rho_p$ is the sphere density, $\lambda_p=E\chi/[(1-2\chi)(1+\chi)]$ and $\mu_p=E/[2(1+\chi)]$ are the Lamé coefficients, $E$ is Young's modulus, $\chi$ is Poisson's ratio, $\xi_p$ is the bulk viscosity and $\eta_p$ is the shear viscosity.
A solution to Eq.~(\ref{eq:109}) is sought as
\begin{equation}
    \boldsymbol{u}=\boldsymbol{\nabla}\varphi_p + \boldsymbol{\nabla}\times\boldsymbol{\psi}_p.
    \label{eq:110}
\end{equation}
Substitution of Eq.~(\ref{eq:110}) into Eq.~(\ref{eq:109}) yields
\begin{align}
    \Delta\varphi_p + k_l^2\varphi_p&=0,\label{eq:111b}\\
    \Delta\boldsymbol{\psi}_p + k_t^2\boldsymbol{\psi}_p&=0,\label{eq:112b}
\end{align}
where $k_l$ and $k_t$ are the longitudinal and the transverse wavenumbers, respectively, which are calculated by
\begin{gather}
    k_l=\omega\sqrt{\frac{\rho_p}{\lambda_p + 2\mu_p - i\omega(\xi_p + 4\eta_p/3)}},\\
    k_t=\omega\sqrt{\frac{\rho_p}{\mu_p - i\omega\eta_p}}.
\end{gather}
Considering Eq.~(\ref{eq:111b}) and Eq.~(\ref{eq:112b}), the wave inside the sphere is written by
\begin{align}
    \varphi_p(r_2,\theta_2,t)&=\mathrm{e}^{-i\omega t}\sum_{n=0}^\infty \hat{a}_nj_n(k_lr_2)P_n(\mu_2),\\
    \psi_p(r_2,\theta_2,t)&=\boldsymbol{e}_{\varepsilon_2}\mathrm{e}^{-i\omega t}\sum_{n=1}^\infty \hat{b}_nj_n(k_tr_2)P_n^1(\mu_2),
\end{align}
leading to the following displacement components:
\begin{align}
    u_r(r_2,\theta_2,t)&=\frac{\mathrm{e}^{-i\omega t}}{r_2}\sum_{n=0}^\infty \big[\hat{a}_nk_lr_2j_n'(k_lr_2) \nonumber\\*
    &- \hat{b}_nn(n+1)j_n(k_tr_2)\big]P_n(\mu_2),\\
    u_\theta(r_2,\theta_2,t)&=\frac{\mathrm{e}^{-i\omega t}}{r_2}\sum_{n=1}^\infty \big\{\hat{a}_nj_n(k_lr_2) \nonumber\\*
    &- \hat{b}_n\left[j_n(k_t r_2) + k_t r_2j_n'(k_tr_2)\right]\big\}P_n^1(\mu_2).
\end{align}
The boundary conditions at the surface of a viscoelastic solid sphere require the continuity of the normal and tangential components of $\boldsymbol{v}$, as well as the continuity of the normal and tangential stress. These conditions can be written by
\begin{gather}
    v_r=\frac{\partial u_r}{\partial t},\label{eq:119}\\
    v_\theta=\frac{\partial u_\theta}{\partial t},\label{eq:120} \\
    \sigma_{r\theta}=\hat{\sigma}_{r\theta},\label{eq:121}\\
    \sigma_{rr}=\hat{\sigma}_{rr},\label{eq:122}
\end{gather}
at $r_2=R_{20}$, and the stress components inside the viscoelastic sphere are calculated by \citep{landau1987fluid, landau1970theory}
\begin{align}
    \hat{\sigma}_{rr}&=\lambda_p\boldsymbol{\nabla}\cdot\boldsymbol{u} + 2\mu_p\frac{\partial u_r}{\partial r_2}\nonumber\\*
    &+2\eta_p\frac{\partial^2 u_r}{\partial t \partial r_2} + \left(\xi_p -\frac{2}{3}\eta_p\right)\boldsymbol{\nabla}\cdot\frac{\partial\boldsymbol{u}}{\partial t}, \\
    \hat{\sigma}_{r\theta}&=\mu_p\left(\frac{1}{r_2}\frac{\partial u_r}{\partial\theta_2}+\frac{\partial u_\theta}{\partial r_2}-\frac{u_\theta}{r_2}\right) \nonumber\\*
    &+ \eta_p\frac{\partial}{\partial t}\left(\frac{1}{r_2}\frac{\partial u_r}{\partial\theta_2}+\frac{\partial u_\theta}{\partial r_2}-\frac{u_\theta}{r_2}\right).
\end{align}
These boundary conditions are calculated in Appendix~\ref{sec:app24}. Combining the resulting equations, given by Eqs.~(\ref{eq:112}) -- (\ref{eq:115}), with the boundary conditions at the surface of the bubble given by Eqs.~(\ref{eq:36}) and (\ref{eq:37}), a system of six equations is obtained, which can be solved numerically in order to find the linear scattering coefficients $a_n^{(1)}$, $a_n^{(2)}$, $b_n^{(1)}$, $b_n^{(2)}$, $\hat{a}_n$ and $\hat{b}_n$.

\subsection{\label{sec:2b} Frequency response of the bubble near a sphere}

In Sec.~\ref{sec:2a}, we derived the equations for the linear scattering coefficients, assuming that the oscillation amplitudes of the bubble (i.e. modal amplitudes $s_n$) were known. In this section, we will assume that the bubble oscillates along all modes, whose amplitudes are unknown. In this case, the perturbed surface of the bubble can be expressed as
\begin{equation}
    R_1(\mu_1,t)=R_{10}+\mathrm{e}^{-i\omega t}\sum_{n=0}^\infty s_nP_n(\mu_1).\label{eq:123}
\end{equation}
As $s_n$ is now considered as unknown, just like the linear scattering coefficients, one more boundary condition is needed to ensure that the number of unknowns matches the number of boundary conditions. Then, one can write the condition for the normal stress on the surface of the bubble, which is given by
\begin{equation}
        P_{g0}\left(\frac{V_{10}}{V_1}\right)^{\gamma}=P_{\mathrm{ac}}\mathrm{e}^{-i\omega t} + P_0-2\sigma H - \sigma_{rr},
    \label{eq:124}
\end{equation}
at $r_1=R_{10}$, where $P_{g0}$ is the equilibrium pressure of the gas inside the bubble, $\gamma$ is the ratio of specific heats of the gas, $V_{10}=4\pi R_{10}^3/3$ is the bubble volume at rest, $V_1$ is the perturbed bubble volume, $P_{\mathrm{ac}}$ is the amplitude of the driving acoustic pressure at the bubble surface, $P_0$ is the hydrostatic pressure in the liquid, $\sigma_{rr}$ is the radial component of the stress tensor in the liquid, $\sigma$ is the surface tension coefficient and $H$ is the mean curvature of the perturbed surface of the bubble. The liquid pressure on the interface, given by Eq.~(\ref{eq:124}), explicitly incorporates the coefficients of the velocity field in the liquid through the normal stress $\sigma_{rr}$ calculated in Eq.~(\ref{eq:72b}).

This additional boundary condition, Eq.~(\ref{eq:124}), is calculated in Appendix~\ref{sec:app25}. The result, given by Eq.~(\ref{eq:136}), is the boundary condition for the normal stress on the surface of the bubble, allowing one to find the oscillation amplitudes $s_n$ as a function of the driving pressure and the other parameters of the problem.

\subsubsection{\label{sec:2b1} Radial pulsation and resonance frequency of the bubble}

The result of Eq.~(\ref{eq:124}) (i.e., Eq.~(\ref{eq:136})) shows that the amplitude of the radial mode $s_0$ is given by
\begin{align}
    s_0&=\frac{1}{\rho_0 R_{10}\omega_0^2}\bigg\{-P_{\mathrm{ac}}+a_0^{(1)}\bigg\{2\eta k_a^2h_0^{(1)''}(x_{a1})\nonumber\\*
    &-k_a^2\bigg[\frac{i\rho_0c^2}{\omega} + \bigg(\xi - \frac{2}{3}\eta\bigg)\bigg]h_0^{(1)}(x_{a1})\bigg\}+\bigg\{2\eta k_a^2j_0''(x_{a1}) \nonumber\\*
    &- k_a^2\bigg[\frac{i\rho_0c^2}{\omega} + \bigg(\xi - \frac{2}{3}\eta\bigg)\bigg]j_0(x_{a1})\bigg\}\sum_{m=0}^\infty a_m^{(2)}F_{0m}\bigg\},\label{eq:62b}
\end{align}
where
\begin{equation}
    \omega_0=\frac{1}{R_{10}}\sqrt{\frac{3\gamma P_{g0}}{\rho_0}-\frac{2\sigma}{\rho_0 R_{10}}}\label{eq:63}
\end{equation}
is the angular resonance frequency of a single bubble with radius $R_{10}$ in an unbounded liquid, $x_{a1}=k_aR_{10}$ and $F_{0m}$ is given by Eq.~(\ref{eq:26}). Assuming that the bubble undergoes only the radial mode and substituting Eq.~(\ref{eq:62b}) into the boundary condition for the normal velocity at the surface of the bubble given by Eq.~(\ref{eq:36}), one obtains
\begin{align}
    &\sum_{m=0}^\infty\left(A_{1nm}^{(1)}+\delta_{m0}\delta_{n0}\alpha_1\right)a_m^{(1)}+\sum_{m=1}^\infty B_{1nm}^{(1)}b_m^{(1)}\nonumber\\*
    +&\sum_{m=0}^\infty\left(A_{1nm}^{(2)}+\delta_{n0}\alpha_2F_{0m}\right)a_m^{(2)}+\sum_{m=1}^\infty B_{1nm}^{(2)}b_m^{(2)}\nonumber\\*
    =&\frac{i\omega P_{\mathrm{ac}}}{\rho_0\omega_0^2}\delta_{n0},\label{eq:64b}
\end{align}
where
\begin{align}
    \alpha_1&=\frac{i\omega}{\rho_0\omega_0^2}\bigg\{2\eta k_a^2h_0^{(1)''}(x_{a1})\nonumber\\*
    &-k_a^2\bigg[\frac{i\rho_0c^2}{\omega} + \bigg(\xi - \frac{2}{3}\eta\bigg)\bigg]h_0^{(1)}(x_{a1})\bigg\},\\
    \alpha_2&=\frac{i\omega}{\rho_0\omega_0^2}\bigg\{2\eta k_a^2j_0''(x_{a1})\nonumber\\*
    &-k_a^2\bigg[\frac{i\rho_0c^2}{\omega} + \bigg(\xi - \frac{2}{3}\eta\bigg)\bigg]j_0(x_{a1})\bigg\},
\end{align}
and $A_{1nm}^{(1)}, B_{1nm}^{(1)}, A_{1nm}^{(2)}$ and $B_{1nm}^{(2)}$ are given by Eqs.~(\ref{eq:40}) -- (\ref{eq:B8}). Solving Eq.~(\ref{eq:64b}) with the other boundary conditions given by Eqs.~(\ref{eq:37}), (\ref{eq:39}) and (\ref{eq:54b}) in the case of a rigid sphere, or with Eqs.~(\ref{eq:37}) and (\ref{eq:81}) -- (\ref{eq:84}) in the case of a fluid sphere, or with Eqs.~(\ref{eq:37}) and (\ref{eq:112}) -- (\ref{eq:115}) in the case of a viscoelastic sphere, we find the scattering coefficients. Substituting the relevant coefficients into Eq.~(\ref{eq:62b}) allows the determination of the radial amplitude $s_0$ as a function of the angular frequency $\omega$ and the distance $d$ to the sphere. Doing so, the resonance frequency can be calculated, taking multiple scattering into account.

\subsubsection{\label{sec:2b2} Shape oscillation amplitude due to multiple scattering}
In addition to the excitation of the radial oscillation of the bubble, the wave scattered by the sphere will excite all modes. It is thus useful to compute the amplitude of the modes $s_n$, and the first boundary condition at the surface of the bubble, Eq.~(\ref{eq:36}), is re-written using Eq.~(\ref{eq:123}), leading to
\begin{align}
    &\sum_{m=0}^{\infty} A_{1 n m}^{(1)} a_m^{(1)}+\sum_{m=1}^{\infty} B_{1 n m}^{(1)} b_m^{(1)}\nonumber\\*
    +&\sum_{m=0}^{\infty} A_{1 n m}^{(2)} a_m^{(2)}+\sum_{m=1}^{\infty} B_{1 n m}^{(2)} b_m^{(2)}\nonumber\\*
    +&i\omega R_{10} s_n=0, \quad n \geq 0.\label{eq:139}
\end{align}
To find the amplitudes of the modes, we can consider $s_n$ as unknown and solve the system of the boundary conditions given by Eqs.~(\ref{eq:136}) and (\ref{eq:139}) along with Eqs.~(\ref{eq:37}), (\ref{eq:39}) and (\ref{eq:54b}) in the case of a rigid sphere, or with Eqs.~(\ref{eq:37}) and (\ref{eq:81}) -- (\ref{eq:84}) in the case of a fluid sphere, or with Eqs.~(\ref{eq:37}) and (\ref{eq:112}) -- (\ref{eq:115}) in the case of a viscoelastic sphere.

An alternative is to express $s_n$ in terms of the scattering coefficients using Eq.~(\ref{eq:139}),
\begin{align}
    s_n&=\frac{i}{\omega R_{10}}\bigg[x_{a1}h_n^{(1)'}(x_{a1}) a_n^{(1)}- n(n+1)h_n^{(1)}(x_{v1})b_n^{(1)}\nonumber\\* 
    &+ x_{a1}j_n'(x_{a1})\sum_{m=0}^\infty F_{nm} a_m^{(2)}\nonumber\\*
    &-n(n+1)j_n(x_{v1})\sum_{m=1}^\infty H_{nm} b_m^{(2)}\bigg], \quad n\geq 0,\label{eq:140}
\end{align}
where $x_{v1}=k_vR_{10}$, and $H_{nm}$ is given by Eq.~(\ref{eq:27}). Substituting Eq.~(\ref{eq:140}) into the boundary condition for the normal stress at the surface of the bubble, given by Eq.~(\ref{eq:136}), the amplitude $s_n$ disappears and only equations in terms of the scattering coefficients remain. Solving these equations and substituting the relevant coefficients in Eq.~(\ref{eq:140}), the modal amplitudes $s_n$ are obtained as a function of $d$ and $\omega$. It is worth noting that this methodology only provides the amplitudes of the modes $s_n$ produced by the scattered wave from the sphere, but not their natural frequencies. 

To find the natural pulsation of the $n$th mode, i.e. the resonance frequency of the $n$th mode near the sphere, another approach is thus required.

\subsubsection{\label{sec:2b3} Resonance frequency of mode $n$}

The external forcing chosen here, of the form $P_{\mathrm{ac}} \mathrm{e}^{-i \omega t}$, describes a uniform pressure field that naturally excites the radial oscillation of the bubble. Due to the multiple scattering induced by the presence of a nearby particle, nonspherical oscillations are triggered at the bubble interface. These oscillations occur at the frequency of the external forcing. Hence, the proposed method does not allow the characterization of the resonance frequency of these modes.

To find the resonance frequencies of the nonspherical modes, an external forcing must be chosen in such a way to excite directly these modes. Considering the problem axisymmetry, a non-uniform pressure field of the form $P_{\mathrm{ac}} \mathrm{e}^{-i \omega t} \sum_{n=0}^{\infty} P_n(\mu_1)$ is chosen. Under such a forcing, the boundary condition for the normal stress at the bubble surface must be revised, leading to
\begin{align}
    P_{g0}\left(\frac{V_{10}}{V_1}\right)^{\gamma}=P_{\mathrm{ac}}\mathrm{e}^{-i\omega t}\sum_{n=0}^\infty P_n(\mu_1) + P_0-2\sigma H - \sigma_{rr},
    \label{eq:124b}
\end{align}
at $r_1=R_{10}$, which, after linearization, gives 
\begin{align}
    &\sum_{m=0}^{\infty} A_{0 n m}^{(1)} a_m^{(1)}+\sum_{m=1}^{\infty} B_{0 n m}^{(1)} b_m^{(1)}\nonumber\\*
    +&\sum_{m=0}^{\infty} A_{0 n m}^{(2)} a_m^{(2)}+\sum_{m=1}^{\infty} B_{0 n m}^{(2)} b_m^{(2)}\nonumber\\*
    +&\sum_{m=0}^{\infty}C_0 s_m=-P_{\mathrm{ac}}R_{10}^2, \quad n \geq 0,\label{eq:144}
\end{align}
where $A_{0nm}^{(1)}, B_{0nm}^{(1)}, A_{0nm}^{(2)}, B_{0nm}^{(2)}$ and  $C_0$ are given by Eqs.~(\ref{eq:B73}) -- (\ref{eq:B77}).
We can now solve the system of boundary conditions given by Eqs.~(\ref{eq:139}) and (\ref{eq:144}) along with Eqs.~(\ref{eq:37}), (\ref{eq:39}) and (\ref{eq:54b}) in the case of a rigid sphere, or with Eqs.~(\ref{eq:37}) and (\ref{eq:81}) -- (\ref{eq:84}) in the case of a fluid sphere, or with Eqs.~(\ref{eq:37}) and (\ref{eq:112}) -- (\ref{eq:115}) in the case of a viscoelastic sphere.

The modal amplitudes $s_n$ are then obtained as a function of the angular frequency $\omega$ and the bubble-sphere distance $d$. Doing so, the angular resonance frequency $\omega_n$ of any shape mode $n$ in the vicinity of the sphere is deduced.
\endgroup

\section{\label{sec:3} Numerical results}
To determine the scattering coefficients $a_n^{(j)}$, $b_n^{(j)}$, $\hat{a}_n$, and $\hat{b}_n$, as well as the oscillation amplitude $s_n$ in the different configurations considered above, an infinite linear system must be solved. Taking advantage that the magnitude of the scattering coefficients decreases with increasing mode number $n$ \citep{doinikov2022acoustic}, the system can be truncated by retaining only the first $N_t$ terms. A convergence analysis is therefore performed for the modal amplitudes $s_n$ for each configuration. This procedure allows identifying an optimal value of $N_t$ that balances computational efficiency with accuracy.

As the current model does not allow the direct calculation of the bubble's resonance frequency, but only the frequency response of the bubble, the resonance frequency is assumed to correspond to the frequency at which the maximum amplitude $s_n$ is reached.

In the following, the liquid surrounding the air bubble and the sphere is assumed to be water. Hence, numerical calculations are performed using the following parameters: $\rho_0=\qty{1000}{\kilo\gram\per\cubic\meter}$, $\eta=\qty{1e-3}{\Pa\s}$, $\xi=\qty{0}{\Pa\s}$, $c=\qty{1500}{\meter\per\s}$, $P_0=\qty{101.3e3}{\Pa}$, $\gamma=1.4$, $\sigma=\qty{0.0727}{\N\per\meter}$, and the radius of the bubble is $R_{10}=\qty{10}{\um}$.

\subsection{\label{sec:3a} Resonance behavior in an unbounded liquid}
As a first verification step, we consider the limiting case where the distance between the bubble and the sphere tends to infinity so that the influence of the sphere becomes negligible.

\subsubsection{\label{sec:3a1} Amplitude and resonance frequency of the radial oscillations}

Assuming purely radial oscillations, the predictions of the present model can be compared with those of the classical linearized Rayleigh–Plesset equation, in which damping is introduced in an \textit{ad hoc} manner to account for viscous and compressibility effects. The amplitude of the linearized radial oscillation of the bubble is given by
\begin{equation}
    s_0^{\mathrm{RP}} = \frac{P_{\mathrm{ac}}}{\rho R_{10} \sqrt{(\omega_0^2 - \omega^2)^2 + \delta^2 \omega^2}},
    \label{eq:57}
\end{equation}
where the damping coefficient $\delta = \delta_\eta + \delta_\mathrm{rad}$ includes the viscous damping, $\delta_\eta = \cfrac{4\eta}{\rho_0 R_{10}^2}$, and the acoustic radiation damping, $\delta_\mathrm{rad} = \cfrac{\omega^2 R_{10}}{c}$ \citep{leighton1994acoustic}. The natural angular frequency $\omega_0$ is defined by Eq.~(\ref{eq:63}).

\begin{figure}[t]
    \centering
    \includegraphics[width=\linewidth]{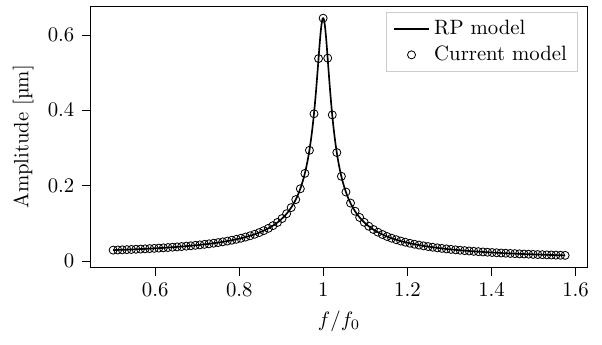}
    \caption{Frequency response of the radial mode for a bubble of radius $R_{10} = \qty{10}{\um}$ driven at $P_{\mathrm{ac}} = \qty{1}{\kilo\Pa}$ in an unbounded liquid. Comparison between the linearized Rayleigh–Plesset model [Eq.~(\ref{eq:57})] and the present model [Eq.~(\ref{eq:59})]. Here, $f_0=\omega_0 / 2\pi$ is the resonance frequency of the bubble in an unbounded liquid, calculated by Eq.~(\ref{eq:63}).}
    \label{fig:FIG2}
\end{figure}

An analogous expression can be derived from the present formulation. The boundary condition at the bubble surface, Eq.~(\ref{eq:30}), is rewritten in the case of a single bubble in an unbounded liquid undergoing purely radial oscillations ($n=0$). The absence of the interaction with the sphere located at infinity means that $a_m^{(2)} = 0$. The first linear scattering coefficient for the wave emitted by the bubble is therefore
\begin{equation}
    a_0^{(1)} = -\frac{i\omega R_{10}}{x_{a1} h_0^{(1)'}(x_{a1})} s_0.
    \label{eq:64}
\end{equation}
Substituting this result into Eq.~(\ref{eq:62b}) for $n=0$ and $a_m^{(2)} = 0$, one obtains
\begin{align}
    s_0 &= -P_{\mathrm{ac}} \Bigg\{ \rho_0 R_{10} \omega_0^2 + \frac{i \omega R_{10}}{x_{a1} h_0^{(1)'}(x_{a1})} \Big[ 2\eta k_a^2 h_0^{(1)''}(x_{a1}) \nonumber \\
    &\quad - k_a^2 \left( \frac{i\rho_0 c^2}{\omega} + \xi - \frac{2}{3}\eta \right) h_0^{(1)}(x_{a1}) \Big] \Bigg\}^{-1}.
    \label{eq:59}
\end{align}
Figure~\ref{fig:FIG2} shows the frequency response of the radial oscillation of a bubble of radius $R_{10}=\qty{10}{\um}$ driven at $P_\text{ac}=\qty{1}{\kilo\Pa}$ in an unbounded liquid, calculated by the classical linearized Rayleigh-Plesset model, Eq.~(\ref{eq:57}), and by the formulation derived in the present paper, Eq.~(\ref{eq:59}). As shown here, the frequency response curves obtained from both models are in excellent agreement, thereby validating the present formulation in the radial case.
\subsubsection{\label{sec:3a2} Resonance frequency of higher-order modes ($n \geq 2$)}

The resonance frequencies associated with non-radial (shape) oscillation modes of a gas bubble were first derived by \citet{lamb1895hydrodynamics} under the assumptions of an incompressible and inviscid liquid. In this idealized case, the natural angular frequency of the $n$th mode is given by
\begin{equation}
    \omega_n = \sqrt{\frac{\sigma (n-1)(n+1)(n+2)}{\rho_0 R_{10}^3}},\quad n\geq 2.
    \label{eq:60}
\end{equation}

\begin{figure}[t]
    \centering
    \includegraphics[width=\linewidth]{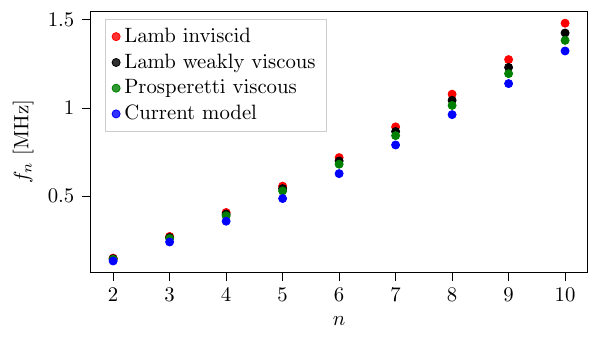}
    \caption{The resonance frequencies of shape modes ($n \geq 2$) for a bubble of radius $R_{10}=\qty{10}{\um}$ in an unbounded liquid. Comparison between the classical Lamb formulation [Eq.~(\ref{eq:60})] for an inviscid and incompressible fluid, Lamb’s weakly viscous approximation, Prosperetti’s viscous and incompressible model and the results provided by the current model (viscous and compressible).}
    \label{fig:FIG3}
\end{figure}

Figure~\ref{fig:FIG3} shows the first ten resonance frequencies of shape modes as predicted by Eq.~(\ref{eq:60}), by various viscous models from the literature, and by the present model, with $d=10^5 R_{10}$ (or equivalently $R_{20}\rightarrow 0$) to simulate the situation where the bubble is in an unbounded liquid. We recall that the external forcing has been modified for investigating shape resonance frequencies, as discussed in Sec.~\ref{sec:2b3}. The results indicate that the predicted resonance frequency given by the present model is systematically lower in comparison to the one predicted by the classical inviscid Lamb formulation, whatever the order $n$ of the shape oscillation. This effect becomes even more pronounced for higher modes $n$. It is well-established that incorporating viscous damping into the physics of a linear oscillator reduces its resonance frequency. Indeed, Lamb also derived a formulation accounting for viscous effects. This weakly viscous model relies on the assumption that the thickness of the viscous penetration layer is small in comparison to the bubble radius. This assumption is generally invalid for a microbubble of radius $R_{10}=\qty{10}{\um}$ driven at frequencies near its resonance. To provide a more rigorous baseline, Fig.~\ref{fig:FIG3} also includes the resonance frequencies derived from the formulation of \citet{prosperetti1977viscous}, which considers the fluid as viscous and incompressible without imposing any restriction on the size of the boundary layer. As expected, Prosperetti's formulation yields lower modal resonance frequencies than Lamb's inviscid formula. Nevertheless, these frequencies remain higher than those computed by the proposed model. This final discrepancy is attributed to the assumption of liquid compressibility, which is known to further lower the resonance frequency. We have checked that removing viscous and compressible effects numerically allows retrieving the same resonance frequencies as the ones given by Eq.~(\ref{eq:60}). To our knowledge, no analytical formulation is currently available in the literature to compute modal resonances in a fluid that is simultaneously viscous and compressible.

\subsection{\label{sec:3b} Resonance behavior of a bubble near a sphere}
In the following, the resonance behavior of both the radial and shape oscillations will be investigated for different types of spheres. To enable comparison with existing models in the literature assuming a radially oscillating bubble near an infinite wall \citep{strasberg1953pulsation, hay2012model, doinikov2015interaction}, spheres with radii up to 40 times the one of the bubble, i.e. up to $R_{20}=\qty{400}{\um}$, will be considered. It is worth noting that a threshold value for the sphere radius exists in the numerical modeling, above which convergence is not reached. This value depends on the physical case (bubble-sphere distance, material of the sphere, oscillation mode). Only cases for which convergence is ensured have been considered for reliability.
\subsubsection{\label{sec:3b1} Rigid sphere}
The case of a rigid sphere is first investigated. A gas bubble with the equilibrium radius $R_{10}=\qty{10}{\um}$ is positioned near a rigid sphere of radius $R_{20}=40R_{10}$, which is located at various distances from the bubble.

\begin{figure}[t]
    \centering
    \includegraphics[width=\linewidth]{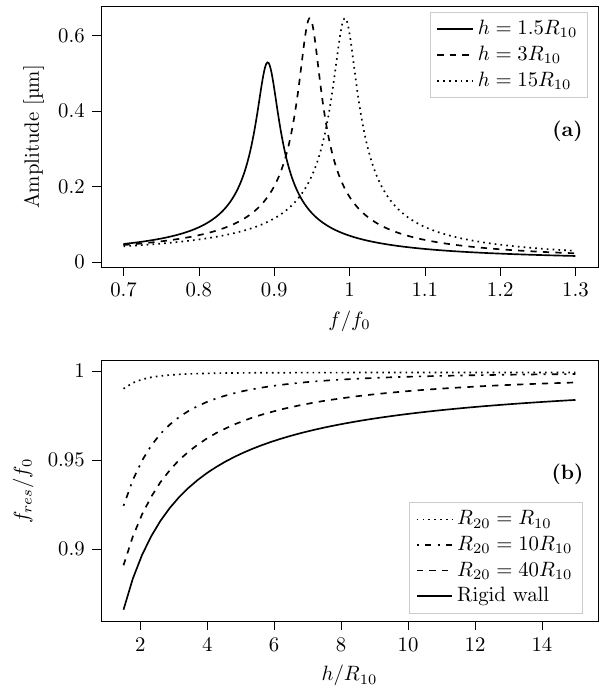}
    \caption{(a) Frequency response of a gas bubble with the equilibrium radius $R_{10} = \qty{10}{\um}$, driven at $P_{\mathrm{ac}} = \qty{1}{\kilo\Pa}$, for various distances $h$ from the surface of a rigid sphere of radius $R_{20} = 40R_{10}$. (b) Normalized resonance frequency of mode $0$ for the same bubble placed at distances $h \in \{1.5R_{10}, \dots, 15R_{10}\}$ from rigid spheres of different radii $R_{20}$.}
    \label{fig:FIG4}
\end{figure}

Figure~\hyperref[fig:FIG4]{\ref*{fig:FIG4}(a)} presents the frequency response of the radial oscillation of the gas bubble driven by a uniform pressure field of amplitude $P_\text{ac}=\qty{1}{\kilo\Pa}$. For the largest distance, $h=15R_{10}$, the resonance curve of the radial oscillation coincides with the one obtained by the linearized Rayleigh-Plesset model in an unbounded fluid, as discussed in Sec.~\ref{sec:3a1}. As the bubble approaches the rigid sphere, both the oscillation amplitude and the resonance frequency of the bubble are clearly reduced. This behavior aligns with experimental observations reported by \citet{garbin2007changes}, who investigated the radial dynamics of a coated microbubble near a rigid planar boundary. Their measurements revealed a marked decrease in the oscillation amplitude as the bubble was positioned closer to the wall, a phenomenon attributed to a downward shift in the resonance frequency due to wall-induced acoustic coupling. The following physical explanation of this decrease can be provided. When a rigid wall is brought close to a pulsating bubble, the fluid motion induced by the bubble must satisfy the no-penetration condition at the wall. This constraint alters the flow field and increases the hydrodynamic interaction between the bubble and its surroundings. In particular, the presence of the wall effectively increases the added mass associated with the bubble’s radial oscillations, leading to a decrease in the resonance frequency. In addition, a larger inertial load opposes the bubble motion, leading to a reduction in the oscillation amplitude for a given acoustic forcing.

The decrease of the resonance frequency of a bubble in front of an infinite rigid wall has been theoretically predicted by \citet{strasberg1953pulsation}. According to the calculation of the total capacitance of two neighboring spheres, Strasberg has derived the resonance frequency shift for a bubble located at a distance $h$ from a rigid wall, given by
\begin{equation}
    \frac{f_{\mathrm{res}}}{f_0}=\left[1+\frac{R_{10}}{2h}\right]^{-1/2},\label{eq:61}
\end{equation}
where $f_0=\omega_0/2\pi$ is the resonance frequency of the bubble in an unbounded liquid, calculated by Eq.~(\ref{eq:63}).

Later, \citet{mobadersany2019acoustic} have extended Strasberg's work to the derivation of the oscillation amplitude of a bubble near a rigid wall. The amplitude of the radial oscillation of a free bubble, given by Eq.~(\ref{eq:57}), is modified by the correction factor $\left(1 + \frac{R_{10}}{2h} \right)^{-1}$ when the bubble is near a wall. This correction leads to a decrease in the radial oscillation amplitude as the wall approaches the bubble.

\begin{figure}[t]
    \centering
    \includegraphics[width=\linewidth]{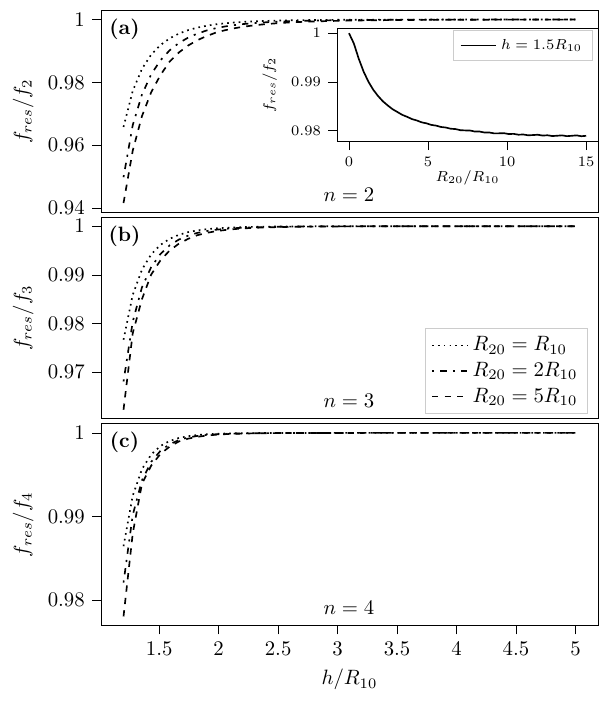}
    \caption{Normalized resonance frequency of shape modes (a) $n = 2$, (b) $n = 3$, and (c) $n = 4$ for a gas bubble with the equilibrium radius $R_{10} = \qty{10}{\micro\meter}$, placed at distances $h \in \{1.2R_{10}, \dots, 5R_{10}\}$ from rigid spheres of varying radius $R_{20}$. The inset in panel (a) shows the normalized resonance frequency of mode $n = 2$ as a function of the sphere size ratio $R_{20}/R_{10}$.}
    \label{fig:FIG5}
\end{figure}

The influence of the rigid sphere on the resonance frequency of the bubble is further investigated in Fig.~\hyperref[fig:FIG4]{\ref*{fig:FIG4}(b)}. The resonance frequency of the bubble near an infinite rigid wall, as given by Eq.~(\ref{eq:61}), is also shown for comparison. The general trend shows a decrease in the bubble resonance frequency with decreasing distances to the sphere, whatever the size of the sphere. The smaller the sphere, the closer it should be in order to influence the bubble resonance frequency. In addition, as the radius of the rigid sphere increases, the resonance frequency curve progressively tends to that of the infinite wall. This highlights the role of the sphere curvature in the bubble dynamics: at the lowest curvature, i.e. for $R_{20}=40R_{10}$, the resonance frequency remains noticeably different from the case of a plane wall (with zero curvature). For experimental applications of bubble-based mechanosensing, this indicates that curvature effects should be taken into account even for relatively large spheres (in comparison to the bubble size). Figure~\hyperref[fig:FIG4]{\ref*{fig:FIG4}(b)} also indicates that the sensitivity of probing materials using bubbles decreases as the size of the probed sphere decreases.

The influence of a nearby sphere on the shape resonance frequencies is shown in Fig.~\ref{fig:FIG5} for the modes with $n=2,3,4$, as a function of the sphere size and the bubble-sphere distance. The resonance frequencies are normalized by the corresponding free-field resonance frequency of each mode allowing for the viscous and compressible effects, computed using the present model and displayed in Fig.~\ref{fig:FIG3}. 

Whatever the considered shape mode, the frequency shift induced by the presence of the sphere occurs only at short distance. Indeed, no noticeable deviation of the shape resonance frequency is observed above $h=2.5 R_{10}$. The resonance frequency of the mode $n=2$ (Fig.~\hyperref[fig:FIG5]{\ref*{fig:FIG5}(a)}) decreases when the bubble comes really close to the sphere, with a noticeable deviation from the shape resonance frequency in an unbounded fluid when $h<2R_{10}$. This effect is more pronounced for large sphere size, but the resonance frequency shift plateaus around 2\% for large $R_{20}$ (see the insert in Fig.~\hyperref[fig:FIG5]{\ref*{fig:FIG5}(a)}). The trend of the variation of the resonance frequency with the bubble-sphere distance is similar for higher modes $n=3,4$ (Fig.~\hyperref[fig:FIG5]{\ref*{fig:FIG5}(b,c)}) but the relative resonance frequency shift decreases with increasing mode number.
\subsubsection{\label{sec:3b2} Fluid sphere}
We now consider the case of an air bubble located near a fluid sphere. As a first step, the sphere is assumed to be air, with the parameters $\rho_s=\qty{1.2}{\kilo\g\per\cubic\meter}$, $c_s=\qty{343}{\meter\per\s}$, $\xi_s=\qty{0}{\Pa\s}$ and $\eta_s=\qty{18.25e-6}{\Pa\s}$.

\begin{figure}[t]
    \centering
    \includegraphics[width=\linewidth]{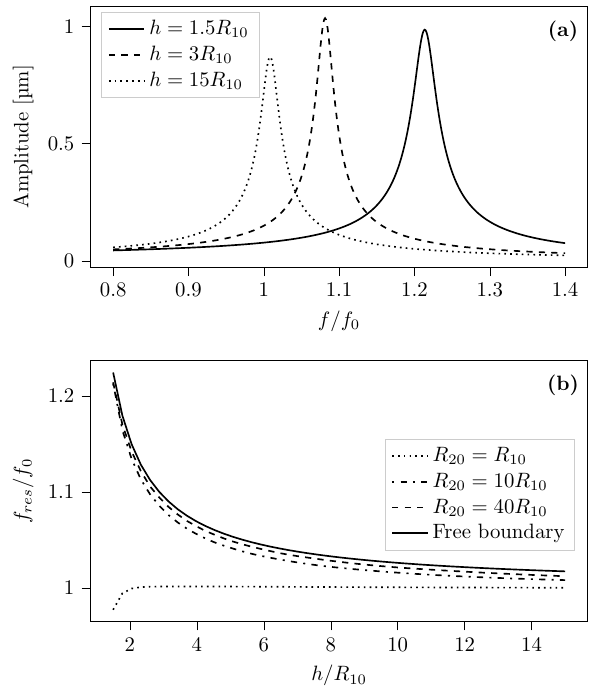}
    \caption{(a) Frequency response of a gas bubble with the equilibrium radius $R_{10} = \qty{10}{\um}$, driven at $P_{\mathrm{ac}} = \qty{1}{\kilo\Pa}$, for various distances $h$ from an air sphere of radius $R_{20} = 10R_{10}$. (b) Normalized resonance frequency of mode $0$ for the same bubble placed at distances $h \in \{1.5R_{10}, \dots, 15R_{10}\}$ from air spheres of different radii $R_{20}$.}
    \label{fig:FIG6}
\end{figure}

The frequency response of a bubble with the equilibrium radius $R_{10} = \qty{10}{\um}$, placed at various distances $h$ from an air sphere of radius $R_{20} = 10R_{10}$, is shown in Fig.~\hyperref[fig:FIG6]{\ref*{fig:FIG6}(a)}. As the bubble approaches the air sphere, both its resonance frequency and its maximum oscillation amplitude increase. Interestingly, the largest oscillation amplitude is observed at the distance $h = 3R_{10}$, while at $h = 1.5R_{10}$, the amplitude is slightly reduced but remains higher than in an unbounded medium. This non-monotonic behavior suggests the existence of an optimal distance that maximizes the acoustic amplitude of the interaction between the bubble and the nearby air interface.

The influence of a free interface on the resonance frequency of a bubble has also been investigated by \citet{strasberg1953pulsation}. His approach, similar to the method used for an infinitely rigid wall, involves introducing a virtual (image) bubble oscillating out of phase with the real bubble. The corresponding resonance frequency is given by
\begin{equation}
    \frac{f_{\mathrm{res}}}{f_0} = \left[1 - \frac{R_{10}}{2h} \right]^{-1/2}, \label{eq:62}
\end{equation}
which corresponds to the resonance frequency of a bubble near a planar boundary with vacuum. 

The influence of the air sphere on the resonance frequency of the bubble is further investigated in Fig.~\hyperref[fig:FIG6]{\ref*{fig:FIG6}(b)}. The case of an infinite free boundary, given by Eq.~(\ref{eq:62}), is plotted as a solid line for comparison. When the bubble and the air sphere have the same radius, the relative resonance frequency remains unchanged until short bubble-sphere distances are reached, here $h<2R_{10}$. This result seems surprising at a first sight, as two equal-sized air cavities oscillating in phase resembles the analogy of a single bubble near a rigid wall with the introduction of the mirror bubble. However, it should be kept in mind that, in the present modeling, the air sphere is assumed to oscillate only in response to the acoustic field radiated by the bubble, without being influenced by the external driving wave (as a consequence, the current analytical framework does not allow one to consider the case of two closely spaced bubbles oscillating in or out of phase). Therefore, the air sphere does not influence the bubble response when located too far, while efficient radiation-induced perturbation of the air sphere starts impacting the bubble dynamics when $h<2R_{10}$. In this case, the analogy with the bubble in front of a wall is recovered, except that the oscillation amplitudes of the two air cavities differ. As a consequence, the relative bubble resonance frequency decreases for short distances. As the radius of the air sphere increases, the relative resonance frequency increases when the bubble-sphere distance decreases, and is higher for larger sphere radius. This behavior converges toward the case of a bubble near an infinite free interface. The agreement between the two models confirms that a sufficiently large compressible sphere can effectively mimic the acoustic response of a planar free boundary.

The influence of a nearby air sphere on the shape resonance frequencies of a bubble has also been investigated. Whatever the bubble or sphere size, bubble-sphere distance, or mode number, the relative resonance frequency shift of a given shape oscillation never exceeds 0.5\% (not shown), being therefore unlikely to be detectable experimentally.

\begin{figure}[t]
    \centering
    \includegraphics[width=\linewidth]{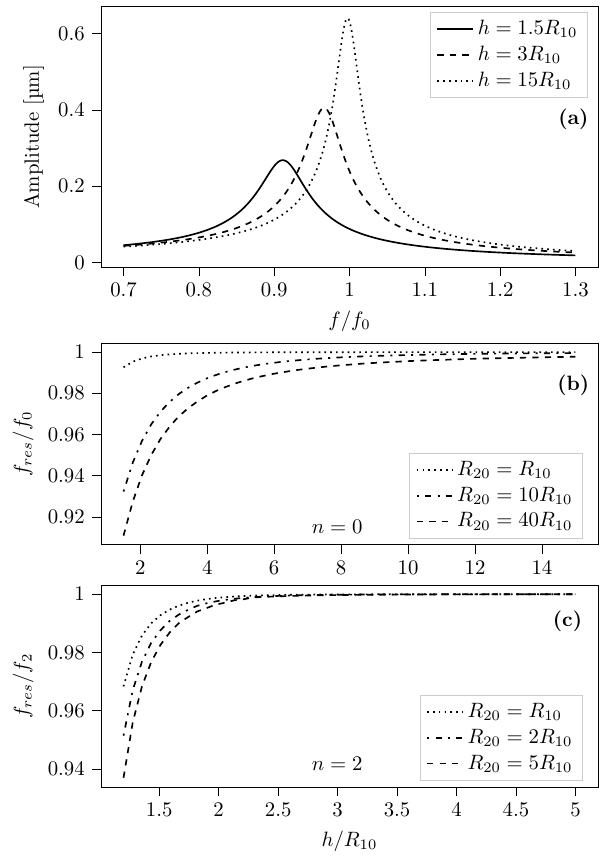}
    \caption{(a) Frequency response of a gas bubble with the equilibrium radius $R_{10} = \qty{10}{\um}$, driven at $P_{\mathrm{ac}} = \qty{1}{\kilo\Pa}$, for various distances $h$ from a glycerin sphere of radius $R_{20} = 40R_{10}$. Normalized resonance frequency of shape modes (b) $n=0$ and (c) $n=2$, for a gas bubble with the equilibrium radius $R_{10}=\qty{10}{\um}$, placed at increasing distances $h$ from a glycerin fluid sphere of varying radius $R_{20}$.}
    \label{fig:FIG7}
\end{figure}

Figure~\ref{fig:FIG7} shows the case of a bubble of radius $R_0=\qty{10}{\um}$ near a glycerin sphere with $\rho_s=\qty{1260}{\kilo\g\per\cubic\meter}$, $c_s=\qty{1920}{\meter\per\s}$, $\xi_s=\qty{0}{\Pa\s}$, and $\eta_s=\qty{1.48}{\Pa\s}$. The frequency response of the bubble at three different distances from the sphere of radii $R_{20}=40R_{10}$ is shown in Fig.~\hyperref[fig:FIG7]{\ref*{fig:FIG7}(a)}. Both the resonance frequency and the radial amplitude decrease with decreasing bubble-sphere distance. The presence of a glycerin sphere near the bubble clearly damps the bubble oscillation, and leads to a nearly 10\% decrease in the relative resonance frequency of the radial oscillation at short distances (Fig.~\hyperref[fig:FIG7]{\ref*{fig:FIG7}(b)}). This significant variation indicates the possibility of probing other materials than purely rigid or soft (air) interfaces using bubbles. Surprisingly, a strong (up to 6\%) variation of the natural frequency of the shape mode $n=2$ is also captured (Fig.~\hyperref[fig:FIG7]{\ref*{fig:FIG7}(c)}), even for relatively small ($R_{20}=5R_{10}$) spherical glycerin inclusions.

\subsubsection{\label{sec:3b3} Viscoelastic sphere}

The nearby sphere is now assumed to be composed of a solid material with viscoelastic mechanical properties. Three materials have been chosen with the aim of investigating various elastic moduli and shear viscosity, namely PMMA, PVA, and CMM whose mechanical properties are given below. When appropriate, the results of the present modeling will be compared with the \citet{strasberg1953pulsation} model of a bubble near a rigid plane, Eq.~(\ref{eq:61}), or with the model of \citet{hay2012model} that predicts the radial dynamics of a bubble close to a viscoelastic wall.

\begin{figure}[t]
    \centering
    \includegraphics[width=\linewidth]{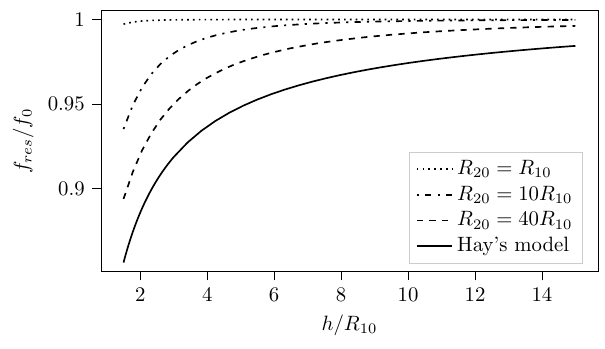}
    \caption{Normalized resonance frequency of the radial mode for a gas bubble with the equilibrium radius $R_{10}=\qty{10}{\um}$, placed at distances $h\in\{1.5R_{10}, \dots, 15R_{10}\}$ from PMMA spheres of varying radius $R_{20}$.}
    \label{fig:FIG8}
\end{figure}

PMMA (polymethyl methacrylate) is an acrylic glass widely used in engineering, whose mechanical properties are $\rho_p=\qty{1170}{\kilo\gram\per\cubic\meter}$, $\xi_p=\qty{0}{\Pa\s}$, $\eta_p=\qty{0}{\Pa\s}$, $E=\qty{3.2}{\giga\Pa}$ and $\chi=0.35$. According to its high elastic modulus, it is expected that the bubble feels the PMMA inclusion as if it were rigid. Indeed, the evolution of its relative resonance frequency with the sphere size and bubble-sphere distance (Fig.~\ref{fig:FIG8}) resembles the case of a rigid sphere (Fig.~\hyperref[fig:FIG4]{\ref*{fig:FIG4}(b)}), with similar trends. When using Hay's model in the case of an infinite plane made of PMMA, one finds that this modeling agrees perfectly with Eq.~(\ref{eq:61}) describing an infinitely rigid material.

\begin{figure}[t]
    \centering
    \includegraphics[width=\linewidth]{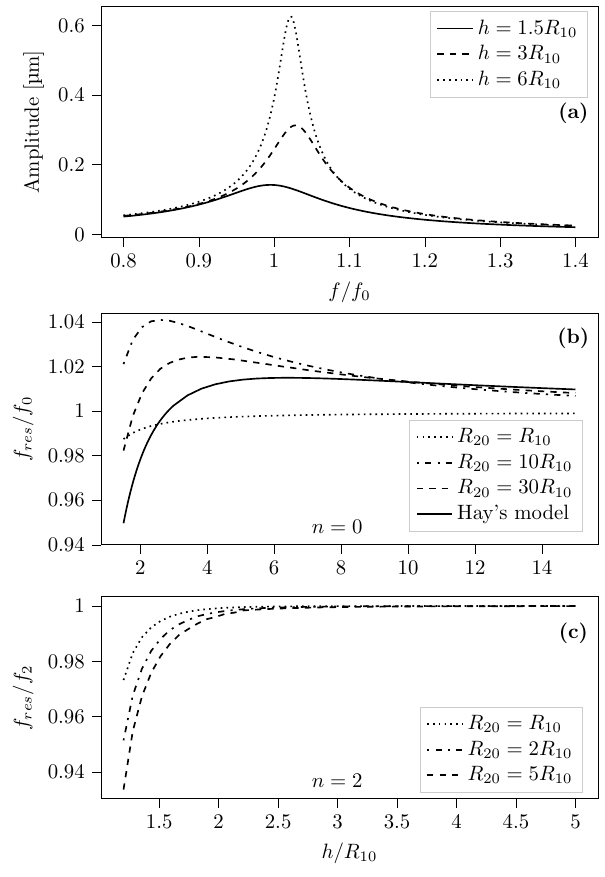}
    \caption{(a) Frequency response of a gas bubble with the equilibrium radius $R_{10} = \qty{10}{\um}$, driven at $P_{\mathrm{ac}} = \qty{1}{\kilo\Pa}$, for various distances $h$ from a PVA sphere of radius $R_{20} = 30R_{10}$. Normalized resonance frequency of oscillation modes (b) $n=0$ and (c) $n=2$, for a gas bubble with the equilibrium radius $R_{10}=\qty{10}{\um}$, placed at increasing distances $h$ from a PVA viscoelastic sphere of varying radius $R_{20}$.}
    \label{fig:FIG9}
\end{figure}

Softer materials can also be considered inside the viscoelastic sphere, such as PVA, which is shown in Fig.~\ref{fig:FIG9}. PVA (polyvinyl alcohol) is a hydrophilic polymer commonly used in biomedical applications, appreciated for its softness and tunable viscoelastic properties. It thus serves as a relevant candidate for modeling soft, tissue-like or biological boundaries. The parameters used in this study are taken from \citet{bisht2024viscoelastic}, corresponding to a specific PVA sample: $\rho_p=\qty{1050}{\kilo\gram\per\cubic\meter}$, $\xi_p=\qty{0}{\Pa\s}$, $\eta_p=\qty{6}{\Pa\s}$, $E=\qty{45}{\kilo\Pa}$, and $\chi=0.495$. In comparison to the PMMA mechanical parameters, the elastic modulus is one hundred times smaller, the shear viscosity is non zero, while other parameters remain more or less identical. 

Figure~\hyperref[fig:FIG9]{\ref*{fig:FIG9}(a)} shows the frequency response of a bubble with the radius $\qty{10}{\um}$ near a PVA sphere of radius $R_{20} = 30R_{10}$, for three different distances $h$ between the bubble and the sphere surface. As the bubble gets closer to the sphere, its oscillation amplitude decreases significantly. A non-monotonous evolution of the resonance frequency as a function of the bubble-sphere distance is observed in Fig.~\hyperref[fig:FIG9]{\ref*{fig:FIG9}(b)}. The resonance frequency first increases when the bubble gets closer to the sphere, reaches a maximum and then decreases at short distances. This behavior is obtained for different sphere radii and tends to the case of an infinite plane modeled by \citet{hay2012model} when the sphere size increases. It can be observed that the resonance frequency predicted by both the present model and Hay's model exhibits a maximum, the position of which depends on the sphere curvature. The case of equal-size bodies $R_{20}=R_{10}$ shows a small variation of the resonance frequency, following the same trend as the one observed in Fig.~\hyperref[fig:FIG6]{\ref*{fig:FIG6}(b)} for an air sphere. It indicates that probing inclusions of the same size as the bubble, regardless of the mechanical properties of the inclusion, is hardly feasible experimentally.

The evolution of the shape resonance frequencies (Fig.~\hyperref[fig:FIG9]{\ref*{fig:FIG9}(c)}) does not exhibit specific signatures of the soft viscoelastic materials and resembles strongly the case of a glycerin sphere (Fig.~\hyperref[fig:FIG7]{\ref*{fig:FIG7}(c)}) with similar trends. Therefore, shape oscillation does not appear to be a good candidate for probing various material elasticity.

\begin{figure}[t]
    \centering
    \includegraphics[width=\linewidth]{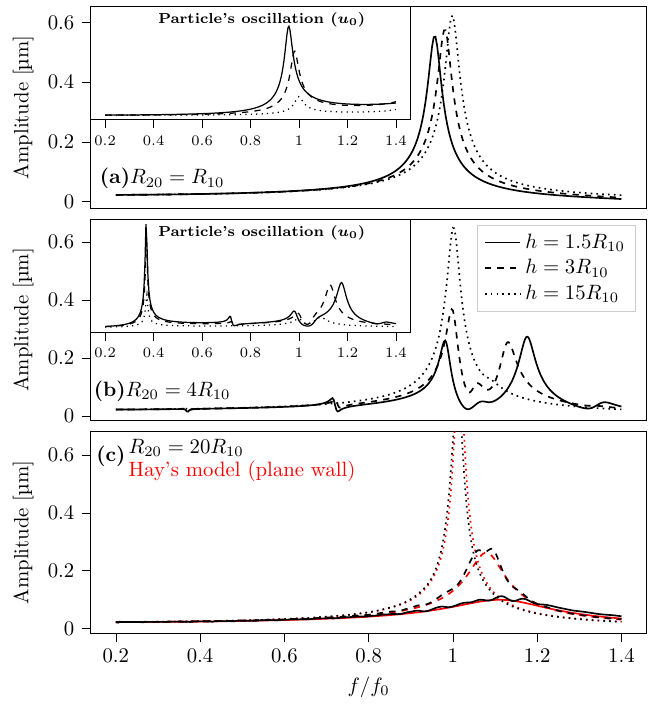}
    \caption{Frequency response of a $\qty{10}{\micro\meter}$-radius bubble driven at $P_{\text{ac}} = \qty{1}{\kilo\pascal}$ at distances $h = \{1.5R_{10}, 3R_{10}, 15R_{10}\}$ from an erythrocyte-like (CMM) particle of radius (a) $R_{20} = R_{10}$, (b) $R_{20} = 4R_{10}$, with the particle's response shown in the inset, and (c) $R_{20} = 20R_{10}$ (black), together with Hay's model prediction for a plane wall with the same viscoelastic parameters (red).}
    \label{fig:FIG10}
\end{figure}

The case of a bubble located near a biological cell is investigated under the material name CMM, for cell-mimicking material. The mechanical parameters of an erythrocyte cell have been selected as $\rho_p = \qty{1125}{\kilo\gram\per\cubic\meter}$, $\xi_p = \qty{0}{\Pa\s}$, $\eta_p = \qty{6}{\milli\Pa\s}$, $E = \qty{26}{\kilo\Pa}$, and $\chi = 0.49$ \cite{dulinska2006stiffness}. The frequency response of a bubble with the radius $R_{10}=\qty{10}{\um}$ near a CMM spherical particle of the same size is shown in Fig.~\hyperref[fig:FIG10]{\ref*{fig:FIG10}(a)}. Both the resonance frequency and the bubble radial oscillation slightly decrease when the bubble comes closer to the cell. When the size ratio between the bubble and the cell increases and reaches $R_{20}=4R_{10}$, multiple resonance peaks are observed on the radial mode of the bubble in Fig.~\hyperref[fig:FIG10]{\ref*{fig:FIG10}(b)}. In fact, the second resonance peak near $f=1.2f_0$ corresponds to one of the acoustic resonances of the cell. To investigate this behavior, we have calculated the frequency response of the particle's oscillations (see the insert in Fig.~\ref{fig:FIG10}). The results reveal that a viscoelastic sphere theoretically exhibits an infinite number of radial resonances for the longitudinal and shear waves. When the resonance frequency of the CMM is close to one of the bubble resonance frequencies, a noticeable interaction occurs between the bubble and the cell. Such a behavior has not been observed with other investigated materials (such as PMMA or PVA) because their first resonances occur at relatively high frequencies. When the CMM-bubble size ratio becomes larger, $R_{20}=20R_{10}$ in Fig.~\hyperref[fig:FIG10]{\ref*{fig:FIG10}(c)}, these secondary peaks vanish, since the particle's resonances no longer overlap with those of the bubble. For this size ratio, an excellent agreement is obtained with the model of \citet{hay2012model} in the case of an infinite wall made of CMM, both for the resonance frequency shift and for the oscillation amplitude. When the bubble gets closer to the large-size cell, its frequency response is strongly attenuated, and its resonance frequency increases with decreasing distances. This reversal in the behavior of the resonance frequency shift indicates that the bubble-cell size ratio is very important when we want to probe material elasticity. One should keep in mind that an actual biological cell is heterogeneous, exhibits a spheroidal shape (instead of spherical) when suspended, and its shape also depends on its adherence properties when attached to a substrate. Predicting the exact sensitivity of such resonances to cell shape or internal inhomogeneities is extremely challenging and a rigorous estimation of these effects would require complex numerical simulations (FEM, etc.) or experimental validations, which are beyond the scope of this linear analytical work.

\subsection{Bubble-mediated mechanosensing}

\begin{figure}[t]
    \centering
    \includegraphics[width=\linewidth]{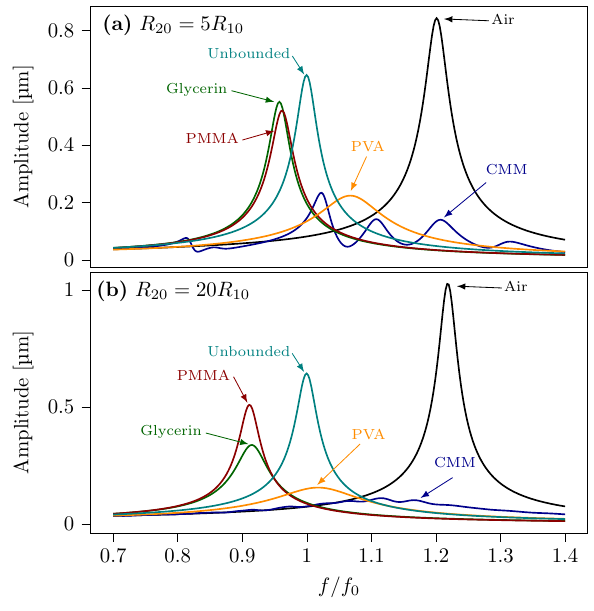}
    \caption{Spectral fingerprint of the bubble near different types of particles, considering a bubble of radius $R_{10}=\qty{10}{\um}$, driven at $P_{ac}=\qty{1}{\kilo\Pa}$, at a distance $h=1.5R_{10}$ from spheres of radius (a) $R_{20}=5R_{10}$ and (b) $R_{20}=20R_{10}$.}
    \label{fig:FIG11}
\end{figure}

The potential of acoustic bubbles to probe the mechanical properties of nearby materials is now examined. A given material will induce a specific modification of the bubble's dynamics, for instance, by modifying its resonance frequency. Such an analysis is comparable with spectroscopic techniques, where different molecular agents are identified by their unique chemical response. Here, the bubble acts as a local acoustic probe, whose resonance is altered by the presence and the nature of a nearby particle. By scanning the frequency response over a range of materials, we obtain a kind of \enquote{acoustic fingerprint} for each particle, opening the door to potential applications in material identification or inverse characterization methods.

The configuration considered here involves a bubble with a $\qty{10}{\um}$-radius, positioned at a fixed distance $h = 1.5 R_{10}$ from spheres of two different sizes, namely $R_{20} = 5 R_{10}$ and $R_{20} = 20 R_{10}$. The spheres are made of different materials with diverse mechanical properties, including Water (unbounded liquid), PMMA, Air, Glycerin, PVA and CMM.

\begin{figure}[t]
    \centering
    \includegraphics[width=\linewidth]{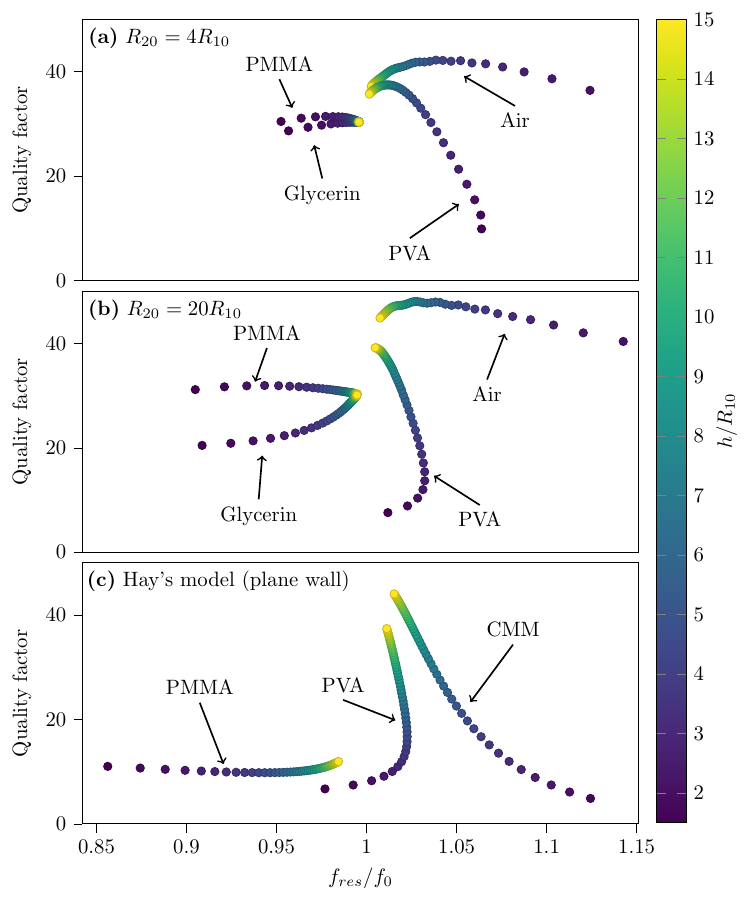}
    \caption{Evolution of the quality factor and the resonance frequency of a bubble of radius $R_{10}=\qty{10}{\um}$ with the distance $h$ between the bubble center and the particle surface for different types of particles of radius (a) $R_{20}=4R_{10}$ and (b) $R_{20}=20R_{10}$. (c) Hay's model prediction for a plane wall.}
    \label{fig:FIG12}
\end{figure}

Figure~\ref{fig:FIG11} shows the bubble spectral fingerprint obtained in this configuration with the materials described above. It can be seen that depending on the material type, size and distance to the sphere, the linear frequency response of the bubble is drastically altered. As expected, a strong mechanical contrast (rigid versus air material) leads to a net modification of the resonance frequency. This indicates that probing rigid or viscoelastic interfaces could be easily performed experimentally. 

However, the frequency response of the bubble appears to be similar for different materials. For instance, no significant differences are observed between glycerin and PMMA, despite their relatively different mechanical properties. This suggests that the analysis of an additional parameter is required to remove such ambiguities. To this end, we propose to analyze the evolution of the quality factor $Q$ of the frequency response curve as a function of the distance to the sphere. 

More precisely, how the bubble can probe each type of material in the three dimensional $(f_{res}/f_0, Q, h/R_{10})$ space is displayed in Fig.~\ref{fig:FIG12}. This choice of representation, where the distance to the bubble is represented with a colormap has emerged as the best possible representation for differentiating materials with very similar resonance properties. Here, the behavior of the resonance frequency shift and the quality factor $Q$ of the resonance curve are calculated for various bubble-sphere distances. Indeed, when scanning a material, the distance of the bubble to the material surface matters and can be perfectly controlled (for instance, when using a caged bubble as in \citet{bouchet2024near}). Specific signatures are obtained for each material in the $(f_{res}/f_0, Q, h/R_{10})$ space. It indicates that adding a third dimension to the scanning (i.e. the bubble-sphere distance) could greatly enhance the accuracy for distinguishing a material from another one. The size ratio also influences the material signature, as in the case of PMMA for which the quality factor $Q$ shows different behavior as a function of the resonance frequency when the sphere size increases from $R_{20}=R_{10}$ to the case of a plane wall. From these results, assessing experimentally the specific spectral signatures of a bubble near an unknown sphere would allow inverting the acoustic problem for retrieving the material properties.

\section{\label{sec:4} Conclusion}
The first-order linearized equations of motion for a viscous compressible liquid have been solved for the specific case of a microbubble oscillating under low-amplitude acoustic excitation near a rigid, liquid, and viscoelastic sphere. The analytical solutions were used to predict the bubble’s dynamic behavior as a function of the applied acoustic signal and the sphere’s physical parameters, focusing in particular on the resonance frequencies of both radial and shape modes. Numerical simulations revealed a strong dependence of the resonance behavior on the material properties and the radius of the sphere. In the limit of large sphere radii, the model converges to existing theoretical models for planar boundaries, thus validating the framework and extending its applicability to interfaces with local curvature. This generalization enables accurate modeling of acoustic interactions between bubbles and small objects such as biological cells. Finally, we propose a scanning-based approach in which the frequency response of the bubble can be used as the spectral signature of a neighboring object, opening the way for mechanical property characterization by means of inverse methods. While this method discriminates materials with high mechanical contrast, its sensitivity to softer biological tissues or specific cell types relies on the accurate mapping of the frequency shift and quality factor at very short interaction distances. Given current experimental constraints at the microscale, this bubble-mediated mechanosensing approach is expected to serve initially as a tool for estimating local mechanical properties.

\begin{acknowledgments}
This work was funded by l'Agence Nationale de la Recherche (ANR), project ANR-22-CE92-0062, and supported by the LabEx CeLyA of the University of Lyon (No. ANR-10-LABX-0060/ANR-11-IDEX-0007).
\end{acknowledgments}

\appendix
\section{Coordinates transformations}
\begingroup
\allowdisplaybreaks
\label{sec:app1}
For the coordinates transformation of $\varphi^{(1)}$ and $\boldsymbol{\psi}^{(1)}$, the following identities are used \citep{FD}:
\begin{align}
h_n^{(1)}\left(k r_1\right) P_n\left(\mu_1\right)&=\frac{1}{(2 n+1) i^n} \sum_{l_1, l_2=0}^{\infty} i^{l_1-l_2}(-1)^{l_2}\nonumber\\*
&\times \left(2 l_1+1\right)\left(2 l_2+1\right)\left(C_{l_1 0 l_2 0}^{n 0}\right)^2\nonumber\\*
&\times h_{l_2}^{(1)}(k d) j_{l_1}\left(k r_2\right) P_{l_1}\left(\mu_2\right),\label{eq:8}\\
h_n^{(1)}\left(k r_1\right) P_n^1\left(\mu_1\right)&=\frac{\sqrt{n(n+1)}}{(2 n+1) i^n} \sum_{l_1,l_2=0}^{\infty} i^{l_1-l_2}(-1)^{l_2}\nonumber\\*
&\times \frac{\left(2 l_1+1\right)\left(2 l_2+1\right)}{\sqrt{l_1\left(l_1+1\right)}} C_{l_1 0 l_2 0}^{n 0} C_{l_1 1 l_2 0}^{n 1}\nonumber\\*
&\times h_{l_2}^{(1)}(k d) j_{l_1}\left(k r_2\right) P_{l_1}^1\left(\mu_2\right),\label{eq:9}
\end{align}
where $C_{l_1m_2l_2m_2}^{LM}$ are the Clebsch-Gordan coefficients \citep{FD, zwillinger2002crc}, and $j_n$ is the spherical Bessel function. Equations~(\ref{eq:8}) and (\ref{eq:9}) follow from Eq.~(34) of \S~5.17 of the book by \citet{FD}.

Substituting Eqs.~(\ref{eq:8}) and (\ref{eq:9}) into Eqs.~(\ref{eq:2}) and (\ref{eq:3}) with $j=1$, wave 1 in the coordinates of the sphere $\left(r_2, \theta_2\right)$ is written by
\begin{align}
\varphi^{(1)}\left(r_2, \theta_2, t\right)&=\mathrm{e}^{-i \omega t} \sum_{n, m=0}^{\infty} a_m^{(1)} E_{n m} j_n\left(k_a r_2\right) P_n\left(\mu_2\right),\label{eq:10} \\
\psi^{(1)}\left(r_2, \theta_2, t\right)&=\mathrm{e}^{-i \omega t} e_{\varepsilon} \sum_{n, m=1}^{\infty} b_m^{(1)} G_{n m} j_n\left(k_v r_2\right) P_n^1\left(\mu_2\right),\label{eq:11}
\end{align}
leading to the following velocity components:
\begin{align}
v_r^{(1)}\left(r_2, \theta_2, t\right)&=\frac{\mathrm{e}^{-i \omega t}}{r_2} \sum_{n, m=0}^{\infty}\left[a_m^{(1)} E_{n m} k_a r_2 j_n'\left(k_a r_2\right)\right.\nonumber\\*
&\left.-b_m^{(1)} n(n+1) G_{n m} j_n\left(k_v r_2\right)\right] P_n\left(\mu_2\right),\label{eq:12} \\
v_\theta^{(1)}\left(r_2, \theta_2, t\right)&=\frac{\mathrm{e}^{-i \omega t}}{r_2} \sum_{\substack{n=1 \\ m=0}}^{\infty}\left\{a_{m}^{(1)} E_{n m} j_n\left(k_a r_2\right)\right.\nonumber\\*
&\left.-b_m^{(1)} G_{n m}\left[j_n\left(k_v r_2\right)+k_v r_2 j_n'\left(k_v r_2\right)\right]\right\}\nonumber\\*
&\times P_n^1\left(\mu_2\right),\label{eq:13}
\end{align}
where
\begin{align}
E_{n m}&=\frac{i^{n-m}(2 n+1)}{2 m+1}\nonumber\\*
&\times\sum_{l=|n-m|}^{n+m}(-i)^{-l}(2 l+1)\left(C_{n 0 l 0}^{m 0}\right)^2 h_l^{(1)}\left(k_a d\right),\label{eq:14} \\
G_{n m}&=\frac{i^{n-m}(2 n+1) \sqrt{m(m+1)}}{(2 m+1) \sqrt{n(n+1)}} \nonumber\\*
&\times\sum_{l=|n-m|}^{n+m}(-i)^{-l}(2 l+1) C_{n 0l0}^{m 0} C_{n1l0}^{m 1} h_l^{(1)}\left(k_v d\right).\label{eq:15}
\end{align}
Note that in Eqs.~(\ref{eq:14}) and (\ref{eq:15}), the triangle inequality has been used, which requires that the indices of $C_{l_1m_1l_2m_2}^{LM}$ satisfy the following conditions: $l_1+l_2-L\geq 0$, $l_1-l_2+L\geq 0$, $-l_1+l_2+L\geq 0$ \citep{zwillinger2002crc}.

The same approach is used for $\varphi^{(2)}$ and $\boldsymbol{\psi}^{(2)}$. With the help of the coordinate transformations

\begin{align}
h_n^{(1)}\left(k r_2\right) P_n\left(\mu_2\right)&=\frac{1}{(2 n+1) i^n} \sum_{l_1, l_2=0}^{\infty} i^{l_1-l_2}\nonumber\\*
&\times\left(2 l_1+1\right)\left(2 l_2+1\right)\left(C_{l_1 0 l_2 0}^{n 0}\right)^2\nonumber\\*
&\times h_{l_2}^{(1)}(k d) j_{l_1}\left(k r_1\right) P_{l_1}\left(\mu_1\right),\label{eq:20}\\
h_n^{(1)}\left(k r_2\right) P_n^1\left(\mu_2\right)&=\frac{\sqrt{n(n+1)}}{(2 n+1) i^n} \sum_{l_1,l_2=0}^{\infty} i^{l_1-l_2}\nonumber\\*
&\times\frac{\left(2 l_1+1\right)\left(2 l_2+1\right)}{\sqrt{l_1\left(l_1+1\right)}} C_{l_1 0 l_2 0}^{n 0} C_{l_1 1 l_2 0}^{n 1}\nonumber\\*
&\times h_{l_2}^{(1)}(k d) j_{l_1}\left(k r_1\right) P_{l_1}^1\left(\mu_1\right),\label{eq:21}
\end{align}
wave 2 in the coordinates of the bubble $\left(r_1, \theta_1\right)$ is given by
\begin{align}
\varphi^{(2)}\left(r_1, \theta_1, t\right)&=\mathrm{e}^{-i \omega t} \sum_{n, m=0}^{\infty} a_m^{(2)} F_{n m} j_n\left(k_a r_1\right) P_n\left(\mu_1\right),\label{eq:22} \\
\boldsymbol{\psi}^{(2)}\left(r_1, \theta_1, t\right)&=\mathrm{e}^{-i \omega t} \boldsymbol{e}_{\varepsilon} \sum_{n, m=1}^{\infty} b_m^{(2)} H_{nm} j_n\left(k_v r_1\right) P_n^1\left(\mu_1\right),\label{eq:23}
\end{align}
leading to the following velocity components:
\begin{align}
v_r^{(2)}\left(r_1, \theta_1, t\right)&=\frac{\mathrm{e}^{-i \omega t}}{r_1} \sum_{n, m=0}^{\infty}\left[a_m^{(2)} F_{nm} k_a r_1 j_n'\left(k_a r_1\right)\right.\nonumber\\*
&\left.-b_m^{(2)} n(n+1) H_{n m} j_n\left(k_v r_1\right)\right] P_n\left(\mu_1\right),\label{eq:24} \\
v_\theta^{(2)}\left(r_1, \theta_1, t\right)&=\frac{\mathrm{e}^{-i \omega t}}{r_1} \sum_{\substack{n=1 \\ m=0}}^{\infty}\left\{a_m^{(2)} F_{nm} j_n\left(k_a r_1\right)\right.\nonumber\\*
&\left.-b_m^{(2)} H_{nm}\left[j_n\left(k_v r_1\right)+k_v r_1 j_n'\left(k_v r_1\right)\right]\right\}\nonumber\\*
&\times P_n^1\left(\mu_1\right),\label{eq:25}
\end{align}
where
\begin{align}
F_{n m}&=\frac{i^{n-m}(2 n+1)}{2 m+1} \nonumber\\*
&\times \sum_{l=|n-m|}^{n+m} i^{-l}(2 l+1)\left(C_{n 0 l 0}^{m 0}\right)^2 h_l^{(1)}\left(k_a d\right),\label{eq:26} \\
H_{n m}&=\frac{i^{n-m}(2 n+1) \sqrt{m(m+1)}}{(2 m+1) \sqrt{n(n+1)}} \nonumber\\*
&\times \sum_{l=|n-m|}^{n+m} i^{-l}(2 l+1) C_{n 0l0}^{m 0} C_{n 1l0}^{m 1} h_l^{(1)}\left(k_v d\right).\label{eq:27}
\end{align}
\section{Boundary conditions}
\label{sec:app2}
\subsection{Bubble -- Normal velocity and tangential stress}
\label{sec:app21}
From Eq.~(\ref{eq:28}), one obtains, for $n\geq 0$,
\begin{align}
    &a_n^{(1)} x_{a1} h_n^{(1) '}\left(x_{a1}\right)-b_n^{(1)} n(n+1) h_n^{(1)}\left(x_{v1}\right)\nonumber\\*
    +&\sum_{m=0}^\infty a_m^{(2)} F_{nm} x_{a1} j_n'\left(x_{a1}\right)\nonumber\\*
    -&\sum_{m=0}^\infty b_m^{(2)} n(n+1) H_{n m} j_n\left(x_{v1}\right)\nonumber\\*
    &=-i\omega R_{10}\sum_{m=M_1}^{M_N}s_m\delta_{mn},\label{eq:30}
\end{align}
where $x_{a1}=k_aR_{10}$, $x_{v1}=k_vR_{10}$ and $\delta_{nm}$ is the Kronecker delta.

From Eq.~(\ref{eq:29}), one obtains, for $n\geq 1$,
\begin{align}
    &2a_n^{(1)} [x_{a1} h_n^{(1)'}(x_{a1})-h_n^{(1)}(x_{a1})]\nonumber\\*
    +& b_n^{(1)} [(2-n^2-n) h_n^{(1)}\left(x_{v1}\right) -x_{v_1}^2 h_n^{(1)''}\left(x_{v1}\right)]\nonumber\\*
    +&\sum_{m=0}^{\infty}2a_m^{(2)} F_{nm} \left[x_{a1} j_n'\left(x_{a1}\right) - j_n(x_{a_1})\right]\nonumber\\*
    +&\sum_{m=0}^\infty b_m^{(2)}H_{n m}[(2-n^2-n) j_n\left(x_{v1}\right) - x_{v1}^2 j_n''\left(x_{v1}\right)]\nonumber\\*
    &=0.\label{eq:31}
\end{align}

Equations~(\ref{eq:30}) and (\ref{eq:31}) can be written in a more convenient way,
\begin{align} &\sum_{m=0}^{\infty} A_{1 n m}^{(1)} a_m^{(1)}+\sum_{m=1}^{\infty} B_{1 n m}^{(1)} b_m^{(1)}\nonumber\\*
+&\sum_{m=0}^{\infty} A_{1 n m}^{(2)} a_m^{(2)}+\sum_{m=1}^{\infty} B_{1 n m}^{(2)} b_m^{(2)}\nonumber\\*
=&-i\omega R_{10}\sum_{m=M_1}^{M_N}s_m\delta_{nm}, \quad n \geq 0, \label{eq:36}\\
&\sum_{m=0}^{\infty} A_{2 n m}^{(1)} a_m^{(1)}+\sum_{m=1}^{\infty} B_{2 n m}^{(1)}b_m^{(1)}\nonumber\\*
+&\sum_{m=0}^{\infty} A_{2 n m}^{(2)} a_m^{(2)}+\sum_{m=1}^{\infty} B_{2 n m}^{(2)} b_m^{(2)}=0, \quad n \geq 1, \label{eq:37}
\end{align}
where
\begin{align}
    A_{1 n m}^{(1)}&=x_{a 1} h_n^{(1)'}(x_{a1})\delta_{nm}, \label{eq:40} \\
    B_{1 n m}^{(1)}&=-n(n+1)h_n^{(1)}(x_{v1})\delta_{nm}, \\
    A_{1 n m}^{(2)}&=x_{a 1} j_n'\left(x_{a 1}\right) F_{n m}, \\ 
    B_{1 n m}^{(2)}&=-n(n+1) j_n\left(x_{v 1}\right) H_{n m}, \label{eq:B8}\\ 
    A_{2 n m}^{(1)}&=2\left[x_{a 1} h_n^{(1)'}\left(x_{a 1}\right)-h_n^{(1)}\left(x_{a 1}\right)\right] \delta_{n m}, \\ 
    B_{2 n m}^{(1)}&=\left[\left(2-n^2-n\right) h_n^{(1)}\left(x_{v 1}\right)-x_{v 1}^2 h_n^{(1)''}\left(x_{v 1}\right)\right] \delta_{n m}, \\ 
    A_{2 n m}^{(2)}&=2\left[x_{a 1} j_n'\left(x_{a 1}\right)-j_n\left(x_{a 1}\right)\right] F_{n m}, \\ 
    B_{2 n m}^{(2)}&=\left[\left(2-n^2-n\right) j_n\left(x_{v 1}\right)-x_{v 1}^2 j_n''\left(x_{v 1}\right)\right] H_{n m}. \label{eq:47}
\end{align}
\subsection{Rigid sphere -- Normal and tangential velocities}
\label{sec:app22}
From Eq.~(\ref{eq:32}), one obtains, for $n\geq 0$,
\begin{align}
    &\sum_{m=0}^{\infty}a_m^{(1)} E_{n m} x_{a2} j_n'\left(x_{a2}\right)\nonumber\\*
    -&\sum_{m=0}^\infty b_m^{(1)} n(n+1) G_{n m} j_n\left(x_{v2}\right)\nonumber\\*
    +&a_n^{(2)} x_{a2} h_n^{(1) '}\left(x_{a2}\right)-b_n^{(2)} n(n+1) h_n^{(1)}\left(x_{v2}\right)=0,\label{eq:34}
\end{align}
where $x_{a2}=k_aR_{20}$ and $x_{v2}=k_vR_{20}$.

From Eq.~(\ref{eq:33}), one obtains, for $n\geq 1$,
\begin{align}
    &\sum_{m=0}^{\infty}a_{m}^{(1)} E_{n m} j_n\left(x_{a2}\right)\nonumber\\*
    +&\sum_{m=0}^{\infty}b_m^{(1)} G_{n m}\left[j_n\left(x_{v2}\right)+x_{v2} j_n'\left(x_{v2}\right)\right]\nonumber\\*
    +&a_n^{(2)} h_n^{(1)}\left(x_{a2}\right)-b_n^{(2)}\left[h_n^{(1)}\left(x_{v2}\right)+x_{v2} h_n^{(1) '}\left(x_{v2}\right)\right]=0.\label{eq:35}
\end{align}
Equations~(\ref{eq:34}) and (\ref{eq:35}) can be written in a more convenient way,
\begin{align}
    &\sum_{m=0}^{\infty} A_{3 n m}^{(1)} a_m^{(1)}+\sum_{m=1}^{\infty} B_{3 n m}^{(1)} b_m^{(1)}\nonumber\\*
    +&\sum_{m=0}^{\infty} A_{3 n m}^{(2)} a_m^{(2)}+\sum_{m=1}^{\infty} B_{3 n m}^{(2)} b_m^{(2)}=0, \quad n \geq 0, \label{eq:39} \\
    &\sum_{m=0}^{\infty} A_{4 n m}^{(1)} a_m^{(1)}+\sum_{m=1}^{\infty} B_{4 n m}^{(1)} b_m^{(1)}\nonumber\\*
    +&\sum_{m=0}^{\infty} A_{4 n m}^{(2)} a_m^{(2)}+\sum_{m=1}^{\infty} B_{4 n m}^{(2)} b_m^{(2)}=0, \quad n \geq 1, \label{eq:54b}
\end{align}
where
\begin{align}
    A_{3 n m}^{(1)}&=x_{a 2} j_n'\left(x_{a 2}\right) E_{n m},\label{eq:B17} \\ 
    B_{3 n m}^{(1)}&=-n(n+1) j_n\left(x_{v 2}\right) G_{n m}, \\ 
    A_{3 n m}^{(2)}&=x_{a 2} h_n^{(1)'}\left(x_{a 2}\right) \delta_{n m}, \\ 
    B_{3 n m}^{(2)}&=-n(n+1) h_n^{(1)}\left(x_{v 2}\right) \delta_{n m}, \\ 
    A_{4 n m}^{(1)}&=j_n\left(x_{a 2}\right) E_{n m}, \\ 
    B_{4 n m}^{(1)}&=-\left[j_n\left(x_{v 2}\right)+x_{v 2} j_n'\left(x_{v 2}\right)\right] G_{n m}, \\ 
    A_{4 n m}^{(2)}&=h_n^{(1)}\left(x_{a 2}\right) \delta_{n m}, \\ 
    B_{4 n m}^{(2)}&=-\left[h_n^{(1)}\left(x_{v 2}\right)+x_{v 2} h_n^{(1)'}\left(x_{v 2}\right)\right] \delta_{n m}.\label{eq:81b}
\end{align}
\subsection{Viscous compressible fluid sphere -- Normal and tangential velocities and stresses}
\label{sec:app23}
From Eq.~(\ref{eq:71}), one obtains, for $n\geq 0$,
\begin{align}
    &x_{a2} j_n'\left(x_{a2}\right)\sum_{m=0}^{\infty}a_m^{(1)} E_{n m}\nonumber\\*
    -&n(n+1)j_n\left(x_{v2}\right) \sum_{m=1}^{\infty}b_m^{(1)} G_{n m}\nonumber \\
    +&a_n^{(2)} x_{a2} h_n^{(1) '}\left(x_{a2}\right)-b_n^{(2)} n(n+1) h_n^{(1)}\left(x_{v2}\right)\nonumber \\
    -&\hat{a}_n\hat{x}_{a2}j_n'(\hat{x}_{a2}) + \hat{b}_nn(n+1)j_n(\hat{x}_{v2}) = 0,
    \label{eq:75}
\end{align}
where $\hat{x}_{a2}=\hat{k}_aR_{20}$ and $\hat{x}_{v2}=\hat{k}_vR_{20}$.

From Eq.~(\ref{eq:72}), one obtains, for $n\geq 1$,
\begin{align}
    &j_n\left(x_{a2}\right)\sum_{m=0}^{\infty}a_{m}^{(1)} E_{n m} \nonumber\\*
    -&\left[j_n\left(x_{v2}\right)+x_{v2} j_n'\left(x_{v2}\right)\right]\sum_{m=1}^{\infty}b_m^{(1)} G_{n m}\nonumber\\*
    +&a_n^{(2)} h_n^{(1)}\left(x_{a2}\right)-b_n^{(2)}\left[h_n^{(1)}\left(x_{v2}\right)+x_{v2} h_n^{(1) '}\left(x_{v2}\right)\right]\nonumber\\*
    -&\hat{a}_nj_n(\hat{x}_{a2}) + \hat{b}_n\left[j_n(\hat{x}_{v2}) + \hat{x}_{v2}j_n'(\hat{x}_{v2})\right]=0.
    \label{eq:76}
\end{align}
From Eq.~(\ref{eq:73}), one obtains, for $n\geq 1$,
\begin{align}
    &2\eta\left[x_{a2} j_n'\left(x_{a2}\right)-j_n(x_{a_2})\right]\sum_{m=0}^{\infty}a_m^{(1)} E_{nm}\nonumber\\*
    +&\eta[(2-n^2-n) j_n\left(x_{v2}\right)- x_{v2}^2 j_n''\left(x_{v2}\right)]\sum_{m=1}^{\infty}b_m^{(1)}G_{n m}\nonumber\\*
    +&2a_n^{(2)}\eta[x_{a2}h_n^{(1)'}(x_{a2})-h_n^{(1)}(x_{a2})] \nonumber\\*
    +&b_n^{(2)}\eta[(2 - n^2 - n)h_n^{(1)}(x_{v2})-x_{v2}^2h_n^{(1)''}(x_{v2})]\nonumber\\*
    -&2\hat{a}_n\eta_s[\hat{x}_{a2}j_n'(\hat{x}_{a2}) - j_n(\hat{x}_{a2})]\nonumber\\*
    -&\hat{b}_n\eta_s[(2 - n^2 - n)j_n(\hat{x}_{v2})-\hat{x}_{v2}^2j_n''(\hat{x}_{v2})]=0.
    \label{eq:77}
\end{align}
From Eq.~(\ref{eq:74}), one obtains, for $n\geq 0$,
\begin{align}
    &x_{a2}^2\left\{2\eta j_n''(x_{a2})-\left[\frac{i\rho_0c^2}{\omega} + \left(\xi - \frac{2}{3}\eta\right)\right]j_n(x_{a2})\right\}\nonumber\\*
    &\times\sum_{m=0}^\infty a_m^{(1)}E_{nm}\nonumber\\*
    +&2\eta n(n+1)[j_n(x_{v2}) - x_{v2}j_n'(x_{v2})]\sum_{m=1}^\infty b_m^{(1)}G_{nm}\nonumber\\*
    +&a_n^{(2)}x_{a2}^2\left\{2\eta h_n^{(1)''}(x_{a2})-\left[\frac{i\rho_0c^2}{\omega} + \left(\xi - \frac{2}{3}\eta\right)\right]h_n^{(1)}(x_{a2})\right\}\nonumber\\*
    +&2\eta b_n^{(2)} n(n+1)[h_n^{(1)}(x_{v2}) - x_{v2}h_n^{(1)'}(x_{v2})]\nonumber\\*
    +&\hat{a}_n\hat{x}_{a2}^2\left\{\left[\frac{i\rho_sc_s^2}{\omega} + \left(\xi_s - \frac{2}{3}\eta_s\right)\right]j_n(\hat{x}_{a2}) - 2\eta_s j_n''(\hat{x}_{a2})\right\}\nonumber\\*
    +&2\eta_s\hat{b}_n n(n+1)[\hat{x}_{v2}j_n'(\hat{x}_{v2})-j_n(\hat{x}_{v2})]=0.
    \label{eq:78}
\end{align}
Equations~(\ref{eq:75}) -- (\ref{eq:78}) can be written in a more convenient way,
\begin{align}
&\sum_{m=0}^{\infty} A_{3 n m}^{(1)} a_m^{(1)}+\sum_{m=1}^{\infty} B_{3 n m}^{(1)} b_m^{(1)}+\sum_{m=0}^{\infty} A_{3 n m}^{(2)} a_m^{(2)}\nonumber\\*
+&\sum_{m=1}^{\infty} B_{3 n m}^{(2)} b_m^{(2)}+\sum_{m=0}^{\infty} \hat{A}_{3 n m} \hat{a}_m+\sum_{m=1}^{\infty} \hat{B}_{3 n m} \hat{b}_m\nonumber\\*
&=0, \quad n \geq 0,\label{eq:81}\\
&\sum_{m=0}^{\infty} A_{4 n m}^{(1)} a_m^{(1)}+\sum_{m=1}^{\infty} B_{4 n m}^{(1)} b_m^{(1)}+\sum_{m=0}^{\infty} A_{4 n m}^{(2)} a_m^{(2)}\nonumber\\*
+&\sum_{m=1}^{\infty} B_{4 n m}^{(2)} b_m^{(2)}+\sum_{m=0}^{\infty} \hat{A}_{4 n m} \hat{a}_m+\sum_{m=1}^{\infty} \hat{B}_{4 n m} \hat{b}_m\nonumber\\*
&=0, \quad n \geq 1,\label{eq:82}\\
&\sum_{m=0}^{\infty} A_{5 n m}^{(1)} a_m^{(1)}+\sum_{m=1}^{\infty} B_{5 n m}^{(1)} b_m^{(1)}+\sum_{m=0}^{\infty} A_{5 n m}^{(2)} a_m^{(2)}\nonumber\\*
+&\sum_{m=1}^{\infty} B_{5 n m}^{(2)} b_m^{(2)}+\sum_{m=0}^{\infty} \hat{A}_{5 n m} \hat{a}_m+\sum_{m=1}^{\infty} \hat{B}_{5 n m} \hat{b}_m\nonumber\\*
&=0, \quad n \geq 1, \label{eq:83}\\
&\sum_{m=0}^{\infty} A_{6 n m}^{(1)} a_m^{(1)}+\sum_{m=1}^{\infty} B_{6 n m}^{(1)} b_m^{(1)}+\sum_{m=0}^{\infty} A_{6 n m}^{(2)} a_m^{(2)}\nonumber\\*
+&\sum_{m=1}^{\infty} B_{6 n m}^{(2)} b_m^{(2)}+\sum_{m=0}^{\infty} \hat{A}_{6 n m} \hat{a}_m+\sum_{m=1}^{\infty} \hat{B}_{6 n m} \hat{b}_m\nonumber\\*
&=0, \quad n \geq 0, \label{eq:84}
\end{align}
where $A_{3}$, $A_{4}$, $B_{3}$ and $B_{4}$ are given by Eqs.~(\ref{eq:B17}) -- (\ref{eq:81b}), while the other quantities are given by 
\begin{align}
    \hat{A}_{3 n m}&=-\hat{x}_{a2}j_n'(\hat{x}_{a2})\delta_{nm},\label{eq:101b}\\
    \hat{B}_{3 n m}&=n(n+1)j_n(\hat{x}_{v2})\delta_{nm},\\
    \hat{A}_{4 n m}&=-j_n(\hat{x}_{a2})\delta_{nm},\\
    \hat{B}_{4 n m}&=[j_n(\hat{x}_{v2}) + \hat{x}_{v2}j_n'(\hat{x}_{v2})]\delta_{nm},\\
    A_{5 n m}^{(1)}&=2\eta\left[x_{a2} j_n'(x_{a2})-j_n(x_{a_2})\right]E_{nm},\label{eq:B37}\\
    B_{5 n m}^{(1)}&=\eta[(2-n^2-n)j_n(x_{v2}) - x_{v2}^2j_n''(x_{v2})]G_{nm},\\
    A_{5 n m}^{(2)}&=2\eta[x_{a2}h_n^{(1)'}-h_n^{(1)}(x_{a2})]\delta_{nm},\\
    B_{5 n m}^{(2)}&=\eta[(2-n^2-n)h_n^{(1)}(x_{v2})-x_{v2}^2h_n^{(1)''}(x_{v2})]\delta_{nm}\label{eq:B40},\\
    \hat{A}_{5 n m}&=-2\eta_s[\hat{x}_{a2}j_n'(\hat{x}_{a2})-j_n(\hat{x}_{a2})]\delta_{nm},\\
    \hat{B}_{5 n m}&=-\eta_s[(2-n^2-n)j_n(\hat{x}_{v2})-\hat{x}_{v2}^2j_n''(\hat{x}_{v2})]\delta_{nm},\\
    A_{6 n m}^{(1)}&=x_{a2}^2\bigg\{2\eta j_n''(x_{a2})\nonumber\\*
    &-\left[\frac{i\rho_0c^2}{\omega} + \left(\xi - \frac{2}{3}\eta\right)\right]j_n(x_{a2})\bigg\}E_{nm},\label{eq:B43}\\
    B_{6 n m}^{(1)}&=2\eta n(n+1)[j_n(x_{v2}) - x_{v2}j_n'(x_{v2})]G_{nm},\\
    A_{6 n m}^{(2)}&=x_{a2}^2\bigg\{2\eta h_n^{(1)''}(x_{a2})\nonumber\\*
    &-\left[\frac{i\rho_0c^2}{\omega} + \left(\xi - \frac{2}{3}\eta\right)\right]h_n^{(1)}(x_{a2})\bigg\}\delta_{nm},\\
    B_{6 n m}^{(2)}&=2\eta n(n+1)[h_n^{(1)}(x_{v2}) - x_{v2}h_n^{(1)'}(x_{v2})]\delta_{nm},\label{eq:B46}\\
    \hat{A}_{6 n m}&=\hat{x}_{a2}^2\bigg\{\left[\frac{i\rho_sc_s^2}{\omega} + \left(\xi_s - \frac{2}{3}\eta_s\right)\right]j_n(\hat{x}_{a2})\nonumber\\*
    &- 2\eta_s j_n''(\hat{x}_{a2})\bigg\}\delta_{nm},\\
    \hat{B}_{6 n m}&=2\eta_s n(n+1)[\hat{x}_{v2}j_n'(\hat{x}_{v2})-j_n(\hat{x}_{v2})]\delta_{nm}.\label{eq:108}
\end{align}

\subsection{Solid viscoelastic sphere -- Normal and tangential velocities and stresses}
\label{sec:app24}
From Eq.~(\ref{eq:119}), one obtains, for $n\geq 0$,
\begin{align}
    &x_{a2} j_n'\left(x_{a2}\right)\sum_{m=0}^{\infty}a_m^{(1)} E_{n m}\nonumber\\*
    -&n(n+1)j_n\left(x_{v2}\right)\sum_{m=1}^{\infty}b_m^{(1)} G_{n m}\nonumber \\*
    +&a_n^{(2)} x_{a2} h_n^{(1) '}\left(x_{a2}\right)-b_n^{(2)} n(n+1) h_n^{(1)}\left(x_{v2}\right)\nonumber \\*
    +&i\omega\hat{a}_nx_lj_n'(x_l) -i\omega \hat{b}_nn(n+1)j_n(x_t)=0,
    \label{eq:125}
\end{align}
where $x_l=k_lR_{20}$ and $x_t=k_tR_{20}$.

From Eq.~(\ref{eq:120}), one obtains, for $n\geq 1$,
\begin{align}
    &j_n\left(x_{a2}\right)\sum_{m=0}^{\infty}a_{m}^{(1)} E_{n m}\nonumber\\*
    -&\left[j_n\left(x_{v2}\right)+x_{v2} j_n'\left(x_{v2}\right)\right]\sum_{m=1}^{\infty}b_m^{(1)} G_{n m}\nonumber\\*
    +&a_n^{(2)} h_n^{(1)}\left(x_{a2}\right)-b_n^{(2)}\left[h_n^{(1)}\left(x_{v2}\right)+x_{v2} h_n^{(1) '}\left(x_{v2}\right)\right]\nonumber\\*
    +&i\omega \hat{a}_nj_n(x_l)-i\omega \hat{b}_n[j_n(x_t) + x_tj_n'(x_t)]=0.
    \label{eq:126}
\end{align}
From Eq.~(\ref{eq:121}), one obtains, for $n\geq 1$,
\begin{align}
    &2\eta\left[x_{a2} j_n'\left(x_{a2}\right)-j_n(x_{a_2})\right]\sum_{m=0}^{\infty}a_m^{(1)} E_{nm} \nonumber\\*
    +&\eta[(2-n^2-n) j_n\left(x_{v2}\right) - x_{v2}^2 j_n''\left(x_{v2}\right)]\sum_{m=1}^{\infty}b_m^{(1)}G_{n m}\nonumber\\*
    +&2a_n^{(2)}\eta[x_{a2}h_n^{(1)'}(x_{a2})-h_n^{(1)}(x_{a2})] \nonumber\\*
    +&b_n^{(2)}\eta[(2 - n^2 - n)h_n^{(1)}(x_{v2})-x_{v2}^2h_n^{(1)''}(x_{v2})]\nonumber\\*
    +&2\hat{a}_n(i\omega\eta_p - \mu_p)[x_lj_n'(x_l) - j_n(x_l)] \nonumber\\*
    +&\hat{b}_n(i\omega\eta_p - \mu_p)[(2-n^2-n)j_n(x_t) - x_t^2j_n''(x_t)]=0.
    \label{eq:127}
\end{align}
From Eq.~(\ref{eq:122}), one obtains, for $n\geq 0$,
\begin{align}
    &x_{a2}^2\left\{2\eta j_n''(x_{a2})-\left[\frac{i\rho_0c^2}{\omega} + \left(\xi - \frac{2}{3}\eta\right)\right]j_n(x_{a2})\right\}\nonumber\\*
    &\times \sum_{m=0}^\infty a_m^{(1)}E_{nm}\nonumber\\*
    +&2\eta n(n+1)[j_n(x_{v2}) - x_{v2}j_n'(x_{v2})]\sum_{m=1}^\infty b_m^{(1)}G_{nm}\nonumber\\*
    +&a_n^{(2)}x_{a2}^2\left\{2\eta h_n^{(1)''}(x_{a2})-\left[\frac{i\rho_0c^2}{\omega} + \left(\xi - \frac{2}{3}\eta\right)\right]h_n^{(1)}(x_{a2})\right\}\nonumber\\*
    +&2\eta b_n^{(2)} n(n+1)[h_n^{(1)}(x_{v2}) - x_{v2}h_n^{(1)'}(x_{v2})]\nonumber\\*
    +&\hat{a}_nx_l^2\left[2(i\omega\eta_p - \mu_p)j_n''(x_l) - \left(i\omega\xi_p - \frac{2}{3}i\omega\eta_p - \lambda_p\right)j_n(x_l)\right]\nonumber\\*
    +&2\hat{b}_n(i\omega\eta_p - \mu_p)n(n+1)[j_n(x_t)-x_tj_n'(x_t)]=0.
    \label{eq:128}
\end{align}
Equations~(\ref{eq:125}) -- (\ref{eq:128}) can be written in a more convenient way,
\begin{align}
&\sum_{m=0}^{\infty} A_{3 n m}^{(1)} a_m^{(1)}+\sum_{m=1}^{\infty} B_{3 n m}^{(1)} b_m^{(1)}+\sum_{m=0}^{\infty} A_{3 n m}^{(2)} a_m^{(2)}\nonumber\\*
+&\sum_{m=1}^{\infty} B_{3 n m}^{(2)} b_m^{(2)}+\sum_{m=0}^{\infty} \hat{A}_{3 n m} \hat{a}_m+\sum_{m=1}^{\infty} \hat{B}_{3 n m} \hat{b}_m\nonumber\\*
&=0, \quad n \geq 0,\label{eq:112}\\
&\sum_{m=0}^{\infty} A_{4 n m}^{(1)} a_m^{(1)}+\sum_{m=1}^{\infty} B_{4 n m}^{(1)} b_m^{(1)}+\sum_{m=0}^{\infty} A_{4 n m}^{(2)} a_m^{(2)}\nonumber\\*
+&\sum_{m=1}^{\infty} B_{4 n m}^{(2)} b_m^{(2)}+\sum_{m=0}^{\infty} \hat{A}_{4 n m} \hat{a}_m+\sum_{m=1}^{\infty} \hat{B}_{4 n m} \hat{b}_m\nonumber\\*
&=0, \quad n \geq 1,\label{eq:113}\\
&\sum_{m=0}^{\infty} A_{5 n m}^{(1)} a_m^{(1)}+\sum_{m=1}^{\infty} B_{5 n m}^{(1)} b_m^{(1)}+\sum_{m=0}^{\infty} A_{5 n m}^{(2)} a_m^{(2)}\nonumber\\*
+&\sum_{m=1}^{\infty} B_{5 n m}^{(2)} b_m^{(2)}+\sum_{m=0}^{\infty} \hat{A}_{5 n m} \hat{a}_m+\sum_{m=1}^{\infty} \hat{B}_{5 n m} \hat{b}_m\nonumber\\*
&=0, \quad n \geq 1, \label{eq:114}\\
&\sum_{m=0}^{\infty} A_{6 n m}^{(1)} a_m^{(1)}+\sum_{m=1}^{\infty} B_{6 n m}^{(1)} b_m^{(1)}+\sum_{m=0}^{\infty} A_{6 n m}^{(2)} a_m^{(2)}\nonumber\\*
+&\sum_{m=1}^{\infty} B_{6 n m}^{(2)} b_m^{(2)}+\sum_{m=0}^{\infty} \hat{A}_{6 n m} \hat{a}_m+\sum_{m=1}^{\infty} \hat{B}_{6 n m} \hat{b}_m\nonumber\\*
&=0, \quad n \geq 0, \label{eq:115}
\end{align}
where $A_{3}$, $A_{4}$, $A_5$, $A_6$, $B_{3}$, $B_{4}$, $B_5$ and $B_6$ are given by Eqs.~(\ref{eq:B17}) -- (\ref{eq:81b}), (\ref{eq:B37}) -- (\ref{eq:B40}) and (\ref{eq:B43}) -- (\ref{eq:B46}), while the other quantities are given by 
\begin{align}
    \hat{A}_{3nm}&=i\omega x_lj_n'(x_l)\delta_{nm},\\
    \hat{B}_{3nm}&=-i\omega n(n+1)j_n(x_t)\delta_{nm},\\
    \hat{A}_{4nm}&=i\omega j_n(x_l)\delta_{nm},\\
    \hat{B}_{4nm}&=-i\omega[j_n(x_t) + x_tj_n'(x_t)]\delta_{nm},\\
    \hat{A}_{5nm}&=2(i\omega\eta_p - \mu_p)[x_lj_n'(x_l)-j_n(x_l)]\delta_{nm},\\
    \hat{B}_{5nm}&=(i\omega\eta_p - \mu_p)[(2-n^2-n)j_n(x_t)-x_t^2j_n''(x_t)]\delta_{nm},\\
    \hat{A}_{6nm}&=x_l^2\bigg[2(i\omega\eta_p - \mu_p)j_n''(x_l) \nonumber\\*
    &- \left(i\omega\xi_p - \frac{2}{3}i\omega\eta_p - \lambda_p\right)j_n(x_l)\bigg]\delta_{nm},\\
    \hat{B}_{6nm}&=2(i\omega\eta_p - \mu_p)n(n+1)[j_n(x_t)-x_tj_n'(x_t)]\delta_{nm}.
\end{align}

\subsection{Bubble -- Normal stress}
\label{sec:app25}
The perturbed bubble volume, accurate to the first-order terms, is calculated by
\begin{align}
    V_1&=\int_0^{2\pi}d\varepsilon_1\int_0^\pi\sin\theta_1 d\theta_1\int_0^{R_{10}+s}r_1^2 dr_1\nonumber\\*
    &\approx V_{10}+4\pi R_{10}^2 s_0\mathrm{e}^{-i\omega t}.
\end{align}
The value of $H$, accurate to the first-order terms, is calculated by
\begin{equation}
    H=-\frac{1}{R_{10}}+\frac{s}{R_{10}^2}+\frac{1}{2R_{10}^2\sin\theta}\frac{\partial}{\partial\theta}\left(\sin\theta\frac{\partial s}{\partial \theta}\right),
\end{equation}
where
\begin{equation}
    s=\mathrm{e}^{-i\omega t}\sum_{n=0}^{\infty}s_nP_n(\mu_1),
\end{equation}
and $H$ can be calculated using the identity
\begin{equation}
    \frac{1}{\sin\theta}\frac{d}{d\theta}\left(\sin\theta\frac{dP_n}{d\theta}\right)=-n(n+1)P_n.
\end{equation}
Linearization of Eq.~(\ref{eq:124}) yields
\begin{equation}
    P_{g0}=P_0+\frac{2\sigma}{R_{10}},
\end{equation}
and
\begin{align}
    &\left.\sigma_{rr}\right\vert_{r_1=R_{10}}=P_{\mathrm{ac}}\mathrm{e}^{-i\omega t} \nonumber\\*
    &+ \mathrm{e}^{-i\omega t}\sum_{n=0}^{\infty}\left[\frac{3\gamma P_{g0}}{R_{10}}\delta_{n0} + \frac{\sigma(n-1)(n+2)}{R_{10}^2}\right]s_nP_n(\mu_1).
    \label{eq:58b}
\end{align}
Calculating $\sigma_{rr}$ at $r_1=R_{10}$ using Eq.~(\ref{eq:72b}) and substituting the results into Eq.~(\ref{eq:58b}) yields, for $n\geq 0$,
\begin{align}
    &x_{a1}^2\left\{\left[\frac{i\rho_0c^2}{\omega} + \left(\xi - \frac{2}{3}\eta\right)\right]h_n^{(1)}(x_{a1}) - 2\eta h_n^{(1)''}(x_{a1})\right\}a_n^{(1)} \nonumber\\*
    +&2n(n+1)\eta\left[x_{v1}h_n^{(1)'}(x_{v1}) - h_n^{(1)}(x_{v1})\right]b_n^{(1)}\nonumber\\*
    +&x_{a1}^2\left\{\left[\frac{i\rho_0c^2}{\omega} + \left(\xi - \frac{2}{3}\eta\right)\right]j_n(x_{a1})-2\eta j_n''(x_{a1})\right\}\nonumber\\*
    &\times \sum_{m=0}^\infty F_{nm}a_m^{(2)}\nonumber\\*
    +&2n(n+1)\eta\left[x_{v1}j_n'(x_{v1})-j_n(x_{v1})\right]\sum_{m=0}^\infty H_{nm}b_m^{(2)}\nonumber\\*
    +&\left[3\gamma P_{g0} R_{10}\delta_{n0}+\sigma (n-1)(n+2)\right]s_n=-P_{\mathrm{ac}}R_{10}^2\delta_{n0}.\label{eq:45b}
\end{align}
This equation can be re-written in the same form as the other boundary conditions,
\begin{align}
    &\sum_{m=0}^{\infty} A_{0 n m}^{(1)} a_m^{(1)}+\sum_{m=1}^{\infty} B_{0 n m}^{(1)} b_m^{(1)}\nonumber\\*
    +&\sum_{m=0}^{\infty} A_{0 n m}^{(2)} a_m^{(2)}+\sum_{m=1}^{\infty} B_{0 n m}^{(2)} b_m^{(2)}+\sum_{m=0}^{\infty}C_0 s_m\nonumber\\*
    =&-P_{\mathrm{ac}}R_{10}^2\delta_{n0}, \quad n \geq 0,\label{eq:136}
\end{align}
where
\begin{align}
    A_{0 n m}^{(1)}&=x_{a1}^2\bigg\{\left[\frac{i\rho_0c^2}{\omega} + \left(\xi - \frac{2}{3}\eta\right)\right]h_n^{(1)}(x_{a1}) \nonumber\\*
    &-2\eta h_n^{(1)''}(x_{a1})\bigg\}\delta_{nm},\label{eq:B73}\\
    B_{0 n m}^{(1)}&=2n(n+1)\eta\left[x_{v1}h_n^{(1)'}(x_{v1}) - h_n^{(1)}(x_{v1})\right]\delta_{nm},\\
    A_{0 n m}^{(2)}&=x_{a1}^2\bigg\{\left[\frac{i\rho_0c^2}{\omega} + \left(\xi - \frac{2}{3}\eta\right)\right]j_n(x_{a1})\nonumber\\*
    &-2\eta j_n''(x_{a1})\bigg\} F_{nm},\\
    B_{0 n m}^{(2)}&=2n(n+1)\eta\left[x_{v1}j_n'(x_{v1})-j_n(x_{v1})\right] H_{nm},\\
    C_{0 n m}&=\left[3\gamma P_{g0} R_{10} \delta_{n0} + \sigma (n-1)(n+2)\right]\delta_{nm}.\label{eq:B77}
\end{align}
\endgroup
\bibliography{REF2_PREStyle}
\end{document}